\documentclass[journal=jacsat,manuscript=article]{achemso}

\usepackage{chemformula} 
\usepackage[T1]{fontenc} 
\usepackage{amssymb}
\usepackage{bm}
\usepackage{amsmath}
\usepackage{accents}
\usepackage{booktabs}
\usepackage{dcolumn}
\usepackage{chemfig}
\usepackage{pdfpages}

\author{Sobin Alosious}
\affiliation{Australian Institute of Bioengineering and Nanotechnology, The University of Queensland, Brisbane, QLD 4072, Australia}
\email{s.alosious@uq.edu.au}
\author{Shern R. Tee}
\affiliation{School of Environment and Science, Griffith University, Nathan, QLD, 4111, Australia}
\affiliation{Australian Institute of Bioengineering and Nanotechnology, The University of Queensland, Brisbane, QLD 4072, Australia}

\author{Debra J. Searles}
\affiliation{Australian Institute of Bioengineering and Nanotechnology, The University of Queensland, Brisbane, QLD 4072, Australia}
\alsoaffiliation{School of Chemistry and Molecular Biosciences, The University of Queensland, Brisbane, QLD 4072, Australia}

\email{d.bernhardt@uq.edu.au}
\title[An \textsf{achemso} demo]
  {Interfacial Thermal Transport and Electrical Performance of Supercapacitors with Graphene/Carbon Nanotube Composite Electrodes }

\abbreviations{IR,NMR,UV}
\keywords{American Chemical Society, \LaTeX}

\begin{document}

%
%
%
%
%

\begin{abstract}

Advanced supercapacitors have great potential to transform how we store and utilize energy, leading to more efficient and sustainable energy systems. This study reveals the structural features influencing the interfacial thermal transport and electrical performances of supercapacitors, using the constant potential and constant charge molecular dynamics simulation techniques. Thermal and electrical properties were calculated for graphene/carbon nanotube composite electrodes and ionic liquid electrolytes with different nanotube diameter, number, layers, and alignments of the nanotubes. The effect of application of a constant potential on the Kapitza resistance is determined for the first time. The vertically aligned CNT structures exhibited higher electrical performance, while the horizontal arrangement showed better thermal performance. Optimum electrode configurations were identified by considering thermal and electrical performance, along with other design factors, such as structural stability, ease of manufacturing, and scalability. After considering all these factors, the horizontally stacked multi-layer CNT arrangement emerged as the optimal electrode structure. The insights gained from this study aid in comprehending the effects of variations in electrode structure, thereby enabling efficient supercapacitor electrode design. 
  
\end{abstract}
\maketitle
\section{INTRODUCTION}

Double layer supercapacitors, also known as ultracapacitors, are energy storage devices that deliver high power and rapid charge-discharge cycles. Compared to batteries, they generally have faster charge and discharge rates, longer lifetimes, and can operate at a broader range of temperatures. However, their energy density tends to be much lower, making research into new supercapacitor materials and structures of great importance. Supercapacitors have numerous applications in various industries, including transportation \cite{jaafar2010sizing} (e.g., electric buses and trains), aerospace \cite{he2021design}  (e.g., satellites), renewable energy \cite{csahin2020hybrid} (e.g., wind turbines and solar cells), and consumer electronics \cite{farhadi2014real} (e.g., smartphones and laptops). They are beneficial for applications that require frequent cycles or continuous operation where traditional batteries may fail. As technology advances, the demand for supercapacitors is expected to increase due to their unique advantages over battery technologies.
\par
The heat generated within the supercapacitor, and its temperature, rise quickly when a large number of supercapacitors are packed in small volumes to attain high energy density. The operating temperature of a supercapacitor has a signiﬁcant influence on its performance characteristics \cite{xiong2015thermal}, as temperature deviations strongly influence the properties of electrolytes (such as viscosity, the solubility of the salt in solvents, ionic conductivity, thermal stability, decomposition, solvent evaporation). Other issues associated with temperature deviations include self-discharge, reduction in capacitance and lifespan  \cite{niu2004comparative,kotz2010aging}.
The heat generation in a supercapacitor is primarily due to irreversible Joule heating, which happens as charging or discharging current passes through the equivalent series resistance (ESR) of the supercapacitor. The ESR includes contributions from the resistance at the interface between the electrode and the current collectors, the electrode and electrolyte resistances, and resistance as ions diffuse through electrode pores or through the separator \cite{xiong2014review,xiong2014graphitic}. Apart from the irreversible Joule heating losses, reversible heat generation caused by the entropic effect occurs while charging or discharging a supercapacitor. Unlike irreversible Joule heating, the reversible heat generation rate is positive during charging and negative during discharging \cite{hijazi2011thermal,guillemet2006multi}. It is essential to dissipate the heat generated in a supercapacitor to maintain the system at a rated temperature, and the interfaces in supercapacitors are the principal barriers to heat dissipation. The electrolyte-electrode interface exhibits the highest heat barrier due to insufficient contact at the solid-liquid interface.
\par
Thermal management of supercapacitors and batteries has been carried out using different cooling techniques \cite{rashidi2022progress}; however, the interfacial thermal resistance, despite being the principal barrier to heat dissipation, has achieved less attention. Thus, extensive knowledge of the molecular-level interfacial thermal transport mechanisms in supercapacitors is critical for optimizing heat dissipation, safety, and stability. Interfacial thermal resistance (ITR), or Kapitza resistance ($R_k$) \cite{pollack1969kapitza}, measures the thermal flow resistance of an interface. In other words, when there is a heat transfer across an interface between two different materials, a temperature discontinuity arises due to the thermal resistance at that interface. This thermal resistance, due to the disparity in the electronic and vibrational characteristics of the two materials in contact, is called interfacial thermal resistance and is given by
\begin{equation}
R_k=\frac{\Delta T}{J_{q}},
\label{eq:Rk}
\end{equation}
where $J_q$ is the heat flux across the interface and $\Delta T$ is the temperature difference between the materials in contact. Computational studies have proven valuable in developing an understanding of the fundamental mechanism of interfacial thermal transport between solid-fluid interfaces \cite{anandakrishnan2023effects,rabani2023enhanced,alosious2020kapitza,alexeev2015kapitza,gonzalez2019implications}, including graphene-ionic liquid interfaces which are more relevant for supercapacitor thermal transport \cite{qian2019ultralow}. Qian \emph{et al.} \cite{qian2018lower} investigated the interfacial thermal resistance and structure at the interface between graphene and ionic liquid and observed that the molecular orientation of ions near the solid surface dominates the degree of heat transfer at the imidazolium-based IL-graphene interface. The increased wettability at the interface can enable thermal transport by improving contact between the IL and graphene, with the alignment of the imidazole ring linearly related to the interfacial thermal resistance. Finally, they have also predicted the lower limit of the ITR ($\sim$6 m$^2$ K GW$^{-1}$) between IL and graphene when imidazole ring is parallel to the graphene sheet. 
\par
Improving a supercapacitor's performance primarily involves maximizing its energy density and capacitance, which can be accomplished through developing new electrode and electrolyte materials \cite{frackowiak2013carbon}. Carbon-based electrodes are widely used in supercapacitors due to their advantages, such as large surface area, good flexibility, high electrical conductivity, good chemical and thermal
stability, and broad potential window. Carbon-based electrode materials in the literature include activated carbon, carbide-derived carbon, carbon monoliths,  and graphene-carbon nanotube composites \cite{barbieri2005capacitance,chmiola2006anomalous,ma2000preparation,garcia2015constant,niu1997high}. P{\'e}an \emph{et al.} \cite{pean2014dynamics} studied the charging of electrified nanoporous carbide-derived carbon (CDC) electrodes with the help of molecular dynamics simulations. Their study revealed that the charging process occurs diffusively, originating from the interface between the CDC and BMI-PF6 electrolyte and gradually penetrating the bulk of the electrode. Similar results were obtained by Maji \emph{et al.} \cite{Maji2021} who carried out experimental and computational studies of an aqueous electrolyte with CDC electrolytes, showing how diffusion rates differ in regions of the supercapacitor. In both cases, the CDC exhibited significant heterogeneities at the nanoscale, comprising a complex pore network with varying pore sizes, setting it apart from other structures like carbon nanotubes and slit pores. Their findings indicated that certain parts of the CDC will start charging earlier than others, despite being located deeper within the electrode, due to the intricate nature of porous networks.
\par
When analyzing the performance of supercapacitors using computer simulation techniques, CNT-graphene electrodes have advantages in that they provide many of the features of realistic systems such as pore networks, different degrees of curvature and a distribution of pore sizes, while being highly reproducible.  Composite electrodes that combine CNTs and graphene possess sufficient intercalation of CNTs between graphene layers, allowing for ample void space for ion diffusion throughout the electrode. Several studies have been conducted to understand the mechanism of CNT-graphene-based electrodes \cite{cheng2011graphene,yang2011design,buglione2012graphene}. McDaniel \cite{mcdaniel2022capacitance} has developed a modified constant potential molecular dynamics algorithm to enable highly eﬃcient
simulations of supercapacitors and predicted the relation between capacitance and structure of five different  graphene/CNT composite electrodes with an ionic liquid electrolyte. They observed that the graphene/CNT contact points behave as "hot spots" with dramatically increased charge separation and reported a 40\% increase in the differential capacitance of a specific electrode model as a result of nanoconfinement effects. Their findings showed how the electrode/electrolyte interface for nanoporous carbon electrodes might exhibit substantial heterogeneity in ion concentration profiles and electrostatic interactions. 
\par
The constant potential molecular dynamics method has been widely used to model the electrodes due to its better accuracy for simulations of electrochemical interfaces. Merlet \emph{et al.} \cite{Merlet2013simulating} compared constant charge (CCM) and constant potential methods (CPM) in modeling supercapacitors based on ionic liquid electrolytes and graphite electrodes and observed a change in fluid structure and time scales over which relaxation phenomena happen, especially for nanoporous supercapacitors. The difference in the structure of the electrolyte increases with an increase in the electrode charge, which may lead to a shift in the thermodynamic properties of the system. They showed that the results obtained from the constant charge method do not accurately represent the system's properties. Furthermore, they found that the heat generation due to the Joule effect during the formation of an electric current follows Ohm's law for the constant potential method, while the constant charge method displays an unphysical temperature jump. However, the direct use of CPM for thermal analysis of electrodes was not possible since electrode particle positions remained immobilized at all times in that implementation of CPM.
\par
In this paper, we investigate the interfacial thermal transport and electrical properties in a graphene-ionic liquid supercapacitor using nonequilibrium molecular dynamics simulations. The interfacial thermal resistance (ITR) calculated using the constant potential and constant charge method is compared for the first time. In addition, we have modeled different types of graphene/CNT composite electrodes by varying geometrical parameters such as diameter, number, layers, and alignment of CNTs. The electrical and thermal performances of each electrode morphology were investigated based on different parameters. Optimum electrode structures with high specific capacitance, improved heat transfer properties, and also considering several design parameters were identified. In what follows, we present a summary of the modeling, the methodology adopted for the simulation, discussions about the obtained results, and finally, some concluding remarks.

\section{METHODOLOGY}
\subsection{Constant Potential Method Overview}

Supercapacitors are often modeled in molecular dynamics (MD) by assigning a uniform and constant charge to all particles in an electrode and observing how the electrolyte's structure and dynamics change with the magnitude of the applied uniform charge.
Such "constant charge" simulations can give reasonable results for structural properties of the electrolyte near a charged electrode but fail in more complicated situations, such as for nanostructured electrodes, high voltages, or complex electrolytes like ionic liquids \cite{Merlet2013simulating, Wang2014}. By contrast, in the constant potential method (CPM) for molecular dynamics (MD) simulations, conductive electrodes are modeled by dynamically updating their partial charges to achieve electrostatic equipotential across each electrode \cite{Siepmann1995}. If the $N_e$ electrode particles are immobile, the partial charges can be efficiently calculated via one additional electrostatic potential evaluation per MD timestep whereas mobile electrodes require an additional $N_e \times N_e$ matrix inversion or an iterative minimization process each timestep. We refer the reader to other publications for the mathematical and implementation details \cite{Scalfi2020}.
CPM MD adds accuracy to electrode-electrolyte simulations by including electrolyte forces due to the induced polarization of the electrode. In particular, CPM MD can realistically simulate nanostructured electrodes which cannot be approximated as uniformly charged planes. Since electrode charges respond dynamically to the electrolyte configuration, charging and discharging dynamics can also be simulated with CPM MD, unlike in constant charge simulations, resulting in physically meaningful equivalent circuit parameters \cite{Pireddu2023}.

\subsection{Evaluation of the Interfacial thermal resistance using Constant Potential and Constant Charge Methods}\label{compare_ITR}

In the first part of our study, we compare CPM to constant charge simulations for evaluating ITR and using it to directly assess the performance of different supercapacitor nanostructures. We assess the effect of different electrode charging procedures in a model supercapacitor (Figure \ref{model}), consisting of planar graphene electrodes with a coarse-grained ionic liquid electrolyte, before analysing more complicated graphene-CNT morphologies in the next part of our study.

In our model supercapacitor, each electrode is made of three layers of graphene, with each layer having different dynamics. For each electrode, the layer furthest from the electrolyte is a base layer and remains immobile, or non-thermalized, throughout the simulation (but still exerts forces on all other electrolyte and electrode atoms), while the next two layers are thermalized and move over the course of the simulation. All the flexible layers are separately modeled using an optimized Tersoff potential \cite{lindsay2010optimized}, while the non-Coulombic interactions between flexible graphene layers and the electrolyte are modeled using Lennard-Jones (LJ) potentials. In addition, the outermost fixed graphene layers on both ends interact with the electrolyte and flexible layers using corresponding LJ parameters.

The Coulombic interaction between electrode and electrolyte is then modelled by adding a lattice of charge sites to the proximal electrode layer closest to the electrolyte, with the same initial positions as the carbon atoms. To enable thermal measurements, we introduce a novel, computationally efficient method where the graphene carbon atoms are fully mobile, but are associated with charge sites that remain fixed to make the CPM calculations faster. This is appropriate in the cases we consider here where the electrode atoms do not move far from their original sites. To see the effect of this approach we treat the charge sites using three different methods to identify the effects of various approximations:

\begin{itemize}
 \item In Method 1, which is our new approach, the charge sites are kept spatially immobile, and their charges are updated using CPM MD with a matrix multiplication (which is computationally efficient but requires immobile charge sites) but the carbon atoms are mobile.
 \item In Method 2, the charge sites are kept spatially immobile, and each site is given a uniform charge that also stays constant over time (positive for the anode and negative for the cathode). The total electrode charge is set to the average value obtained from the Method 1 simulations, but constant potential is not used to update individual electrode charges. The carbon atoms are mobile.
 \item In Method 3, uniform, constant charges are still used for the charge sites, but they are now made mobile and move in tandem with the carbon atom positions used for non-Coulombic interactions, whereas the previous two methods use static charge sites and mobile non-Coulombic sites.
\end{itemize}

Thus, the difference between results from Method 1 and Method 2 is due to changing from CPM MD to constant charge MD, while the difference between results from Method 2 and Method 3 is due to changing from immobile to thermally-mobile fixed-charge sites.

\subsection{Molecular Dynamics Simulation Details}

\begin{figure}[t!]
\centering
\includegraphics[width=0.85\textwidth]{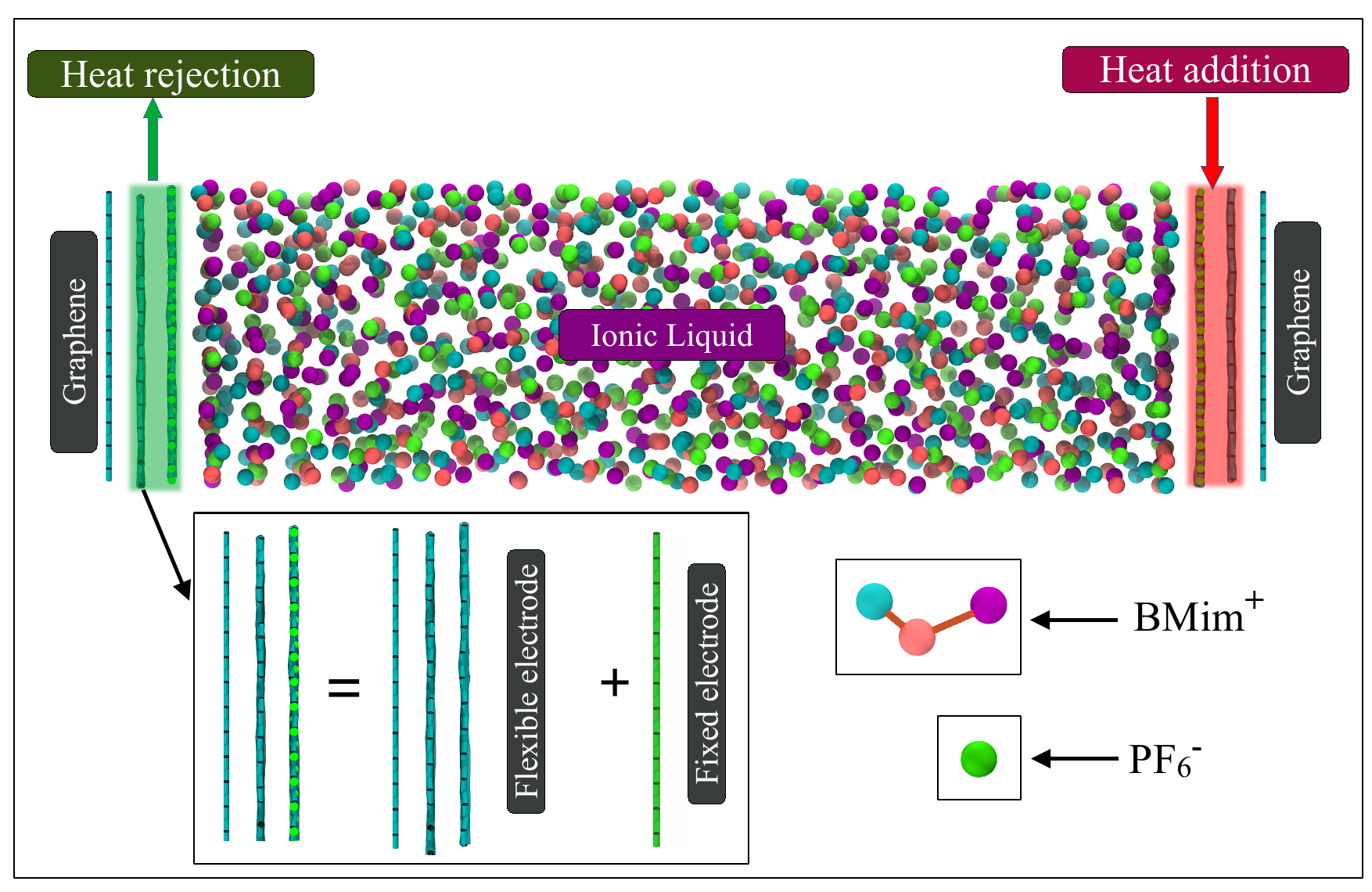}
\caption{Schematic depiction of the graphene-ionic liquid supercapacitor model. The electrode is divided into a flexible and fixed graphene layer for thermal and electrical analysis. }
\label{model}
\end{figure}

The supercapacitor considered for the initial study consists of an ionic liquid electrolyte confined between planar graphene electrodes, schematically depicted in Fig. \ref{model}. Both electrodes consist of three layers of graphene sheets separated by a distance of 3.35 \AA \cite{slack1962anisotropic}. The intermolecular interactions of the IL are via a coarse-grained model of 1-n-butyl-3-methylimidazolium hexafluorophosphate [BMim$^+$ ][PF$_6^-$ ] ionic liquid developed by Roy and Maroncelli \cite{roy2010improved}. In this model, the cations are represented as three-site molecules that are kept rigid using the SHAKE \cite{ciccotti1982molecular} algorithm, while the anions are represented as single particles. Several bulk characteristics, including molar volume, isothermal compressibility, viscosity, and diffusion, can be reliably predicted using this IL model. Periodic boundary conditions were imposed in the \textit{x}, and \textit{y} directions  with dimensions 32.2 \AA, and 34.4 \AA, respectively. The distance between the proximal electrodes is kept at 109.75 \AA\ to maintain the bulk density of the IL (1.267 g cm$^{-3}$) at the center bulk portion of the cell. 
Molecular dynamics simulation models of supercapacitors with similar electrode-electrolyte combinations, force fields, and dimensions have been used in previous works \cite{Merlet2011imidazolium,tee2022fully}. The long-range electrostatic forces were estimated by employing the particle-particle-particle-mesh (PPPM) \cite{hockney2021computer} solver with an accuracy of 10$^{-8}$, and the Ewald summation technique with an extended volume ratio of 3.0 was used to adjust the \textit{z}-direction confinement \cite{yeh1999ewald}.
The pairwise interactions between all the atoms were calculated using LJ and Coulombic potentials with a cut-off distance of 16 \AA. The Lorentz-Berthelot mixing rules were used to obtain the LJ parameters for the cross-terms.
The equations of motion of all the particles were integrated using a velocity-Verlet algorithm with a time step of 1 fs. The Large-scale Atomic/Molecular Massively Parallel Simulator package (LAMMPS) \cite{plimpton1995fast} was used for conducting all the MD simulations, and the ELECTRODE \cite{siepmann1995influence} package was utilized for the Constant Potential Method. The visualization and rendering of models were carried out using Visual Molecular Dynamics (VMD) \cite{humphrey1996vmd} and Open Visualization Tool (OVITO) \cite{stukowski2009visualization}.

The initial modeling of the simulation setup was carried out by following the similar steps as in our previous work, and the same model is used in this study as well \cite{tee2022fully}. The system is equilibrated, and the electrode distance is adjusted to obtain the bulk density of 1.267 g cm$^{-3}$ at the center of the IL channel. Charge sites are then added to the system as needed (for Methods 1 and 2), and the entire system is equilibrated under a canonical (NVT) ensemble at 400 K for a time period of 5.0 ns after energy minimization. During this process, a thermostat is applied to the flexible electrode and the IL, and one of the three charge methods in the previous section is used. To examine how potential differences affect interfacial thermal resistance, nine different potential differences (0.0, 0.5, 1.0, 1.5, 2.0, 2.5, 3.0, 4.0, and 5.0 V) were utilized. As used in previous studies \cite{wang2014evaluation}, the electrode charges were modeled as Gaussian distributions with a width parameter $ \eta $ = 1.979 \AA.\cite{tee2022fully} After this equilibration process, the system will be at a constant temperature, and the average charge on the electrodes will have converged to constant values (for Method 1 simulations). After the equilibration process, NEMD simulations are carried out to calculate the interfacial thermal resistance at the electrode-electrolyte interface. To generate a temperature gradient across the system, heat is added and removed from both ends of the system using a Langevin thermostat with a damping parameter of 100 fs. \cite{grest1986molecular}. A  linear temperature gradient along the \textit{z}-direction is generated by providing two different temperatures (350 K and 450 K) on both ends of the system, as shown in Fig. \ref{model}. This process was continued for 5.0 ns until the system reached a steady state, at which point the net heat flux across the system would converge to a constant value. The ITR can be calculated using the Eq. (\ref{eq:Rk}). The temperature drop at the interface is measured from the temperature profile, and the heat flux is calculated using the energy added and subtracted from the thermostat per unit surface area per unit time. Another set of simulations was conducted to study the change in ITR during the charging process. To calculate the ITR variation during the charging process, the constant potential method is turned off during the initial equilibration and heat addition/removal process and later turned on during the production run. Thus, the charging process begins at the production run, which allows us to calculate the ITR during specific intervals of the charging process.

\subsection{Graphene/CNT Composite Electrodes}
\begin{figure}[b!]
\centering
\includegraphics[width=.9\textwidth]{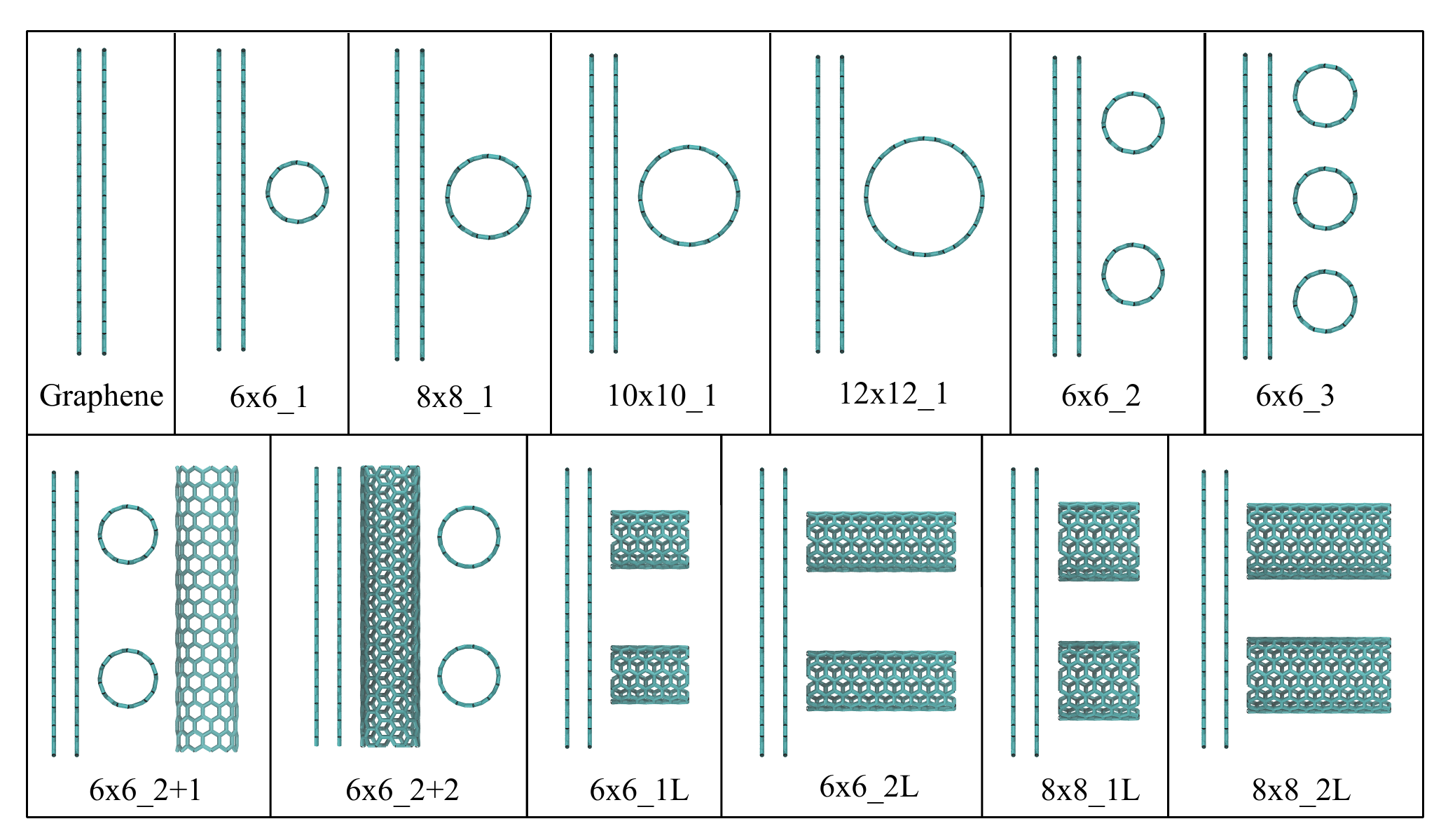}
\caption{The different morphologies of graphene/CNT composite electrodes considered here are classified based on the diameter, number, layers, and alignment of CNTs. The notation used for electrodes is indicated and is based on the characteristics of CNTs. }
\label{electrode}
\end{figure}
The second part of this study involved analyzing the electrical and thermal performance of different graphene/CNT composite electrode morphologies. Graphene/CNT composite electrodes with 14 different combinations are used in this study. The electrode arrangements can be classified into four categories depending upon the CNT diameter, number of CNTs, number of layers of CNT, and the arrangement of CNTs (vertical/horizontal). A typical electrode in this study comprises a graphene layer with \textit{Lx} = \textit{Ly} = 40 \AA\ and different carbon nanotube (CNT) combinations. All the electrode structures and their denotations mentioned throughout this paper are shown in Fig. \ref{electrode}. The effect of the CNT diameter was studied by using four different diameters varying from 8.25 \AA\ to 16.5 \AA\ ((6,6), (8,8), (10,10), and (12,12)). A CNT of 40 \AA\ length is placed at the center of the graphene layer, with its axis parallel to the \textit{x} direction. The graphene layers are periodic in the \textit{x} and \textit{y} directions, whereas the CNT is periodic in the \textit{x} direction. Similarly, the number of CNTs on the electrode varied from 1 to 3 using (6,6) CNTs placed equidistant and parallel to the graphene layer. The complexity and porosity of the electrode are again increased by adding another layer of CNT on top of the first layer. Two such models are prepared by varying the number of CNTs on the second layer. The second layer is also a (6,6) CNT with a length of 40 \AA\ and its axis parallel to the \textit{y} direction. 
\begin{figure}[b!]
\centering
\includegraphics[width=0.85\textwidth]{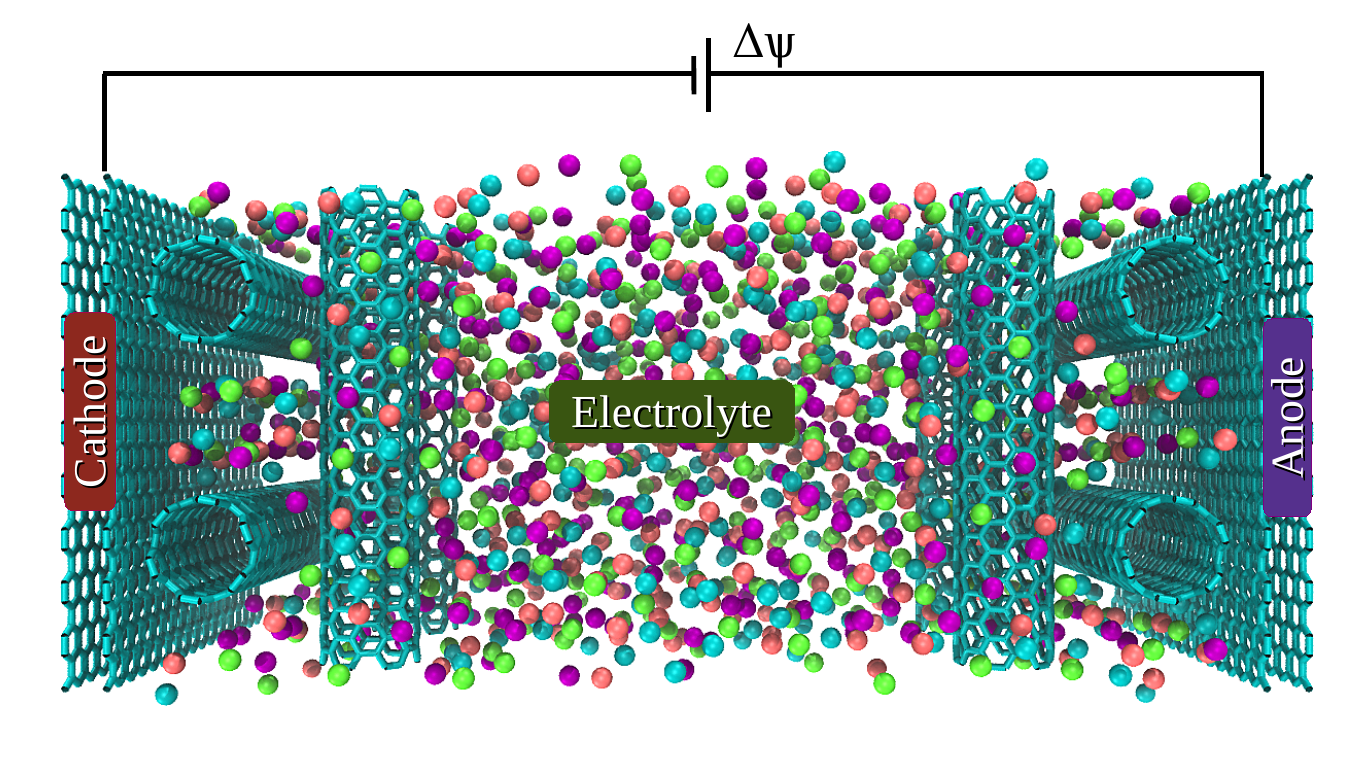}
\caption{Schematic diagram of the two-layer (6x6\_2+2) supercapacitor model. The pores of the electrode are filled by the ions by placing the system under 1 bar pressure at 400 K.  }
\label{sample}
\end{figure}
For initialization, the electrodes in this study were positioned with a packing distance of 3.35 \AA\ between graphene-CNT and CNT-CNT. All the individual atoms of CNTs are tethered to their initial positions using a weak harmonic spring potential, ensuring the structure of the electrodes remains the same. Finally, another four sets of electrodes were modeled by placing the CNTs perpendicular to the graphene. Four CNTs were positioned with their axes perpendicular to the graphene plane (parallel to the \textit{z}-direction). Varying the length and diameter of the CNTs results in four distinct configurations, utilizing a combination of (6,6) and (8,8) CNTs with lengths of 10 \AA\ and 20 \AA\. The supercapacitor models are generated by filling [BMim$^+$ ][PF$_6^-$ ] ionic liquid between the symmetric anode and cathode configurations corresponding to each electrode. The distance between the anode and cathode in each system is adjusted to maintain constant pressure and temperature. This is achieved by applying a force equivalent to 1 bar of pressure on top of one electrode (the other electrode is kept fixed) and allowing it to equilibrate until the pressure stabilizes. Once the system is stabilized, the average distance between the cathode and anode is calculated and set as the distance between electrodes. This process is repeated for all systems to ensure proper electrolyte filling inside the pores of the graphene/CNT composite electrodes. 

During each production run, the electrodes are modeled using the constant potential method with flexible carbon atoms (using the same modified Tersoff potential as above) and immobile charge sites (Method 1). Again, the outermost graphene layers on both sides are always kept immobile. A schematic diagram of the supercapacitor used in this study, featuring a graphene/CNT composite electrode and IL electrolyte, is shown in Fig. \ref{sample}. This particular system consists of two layers of (6,6) CNTs stacked on top of each other to form a two-layer CNT arrangement. The potential difference across the electrodes was varied from 0.0 to 4.0 V to study supercapacitors' thermal and electrical properties under varying potential differences. Finally, the NEMD simulation steps mentioned in the previous section is followed for the thermal analysis.

\section{RESULTS AND DISCUSSIONS}
\subsection{Comparison of Constant Potential and Constant Charge Method}
\label{compare_CPM_CCM}
\begin{figure}[h!]
\centering
\includegraphics[width=0.85\textwidth]{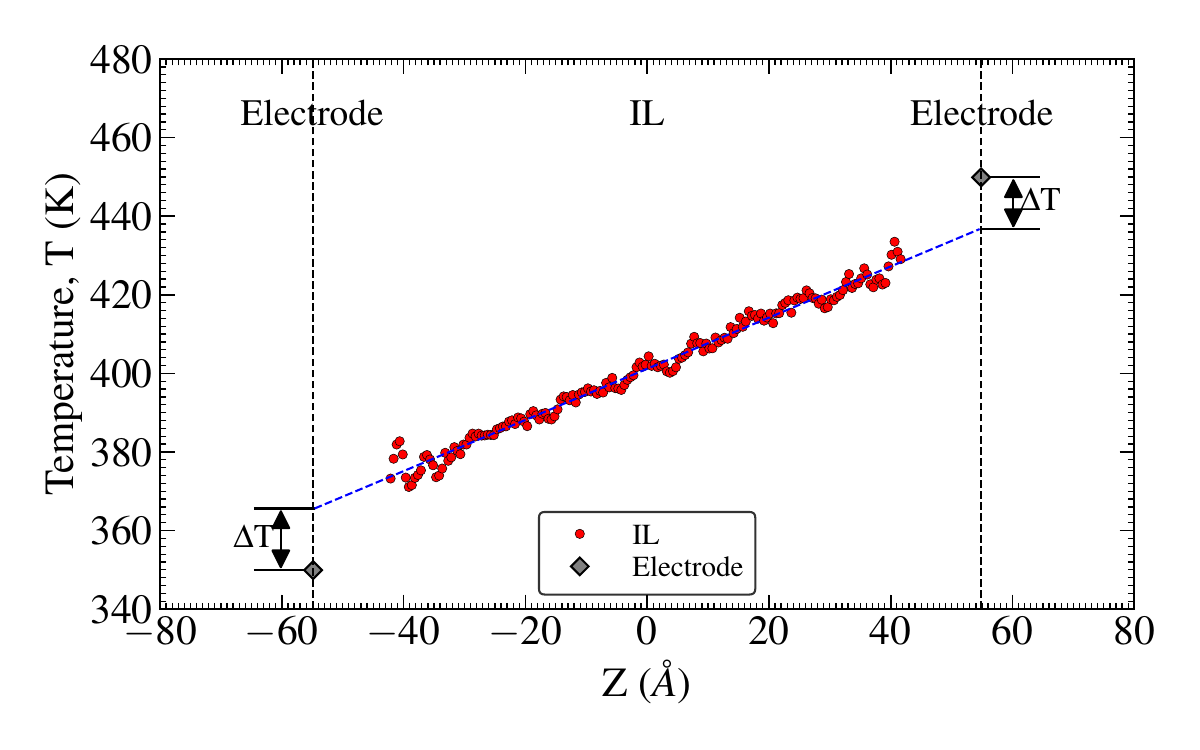}
\caption{Temperature profile along the confinement direction of the graphene supercapacitor.  }
\label{temprofile}
\end{figure}

The interfacial thermal transport is investigated in a graphene-IL supercapacitor by conducting nonequilibrium molecular dynamics simulations, where a temperature gradient along the \textit{z} direction is generated by adding and removing heat from the anode and cathode, respectively. The temperature profile across the system after reaching a steady-state heat flux is illustrated in Fig. \ref{temprofile}. A linear temperature gradient is apparent across the IL channel. However, there is a temperature discontinuity at the interface between the electrode and electrolyte. This discontinuity or jump in temperature, $\Delta T$, is due to the interfacial thermal resistance between graphene and IL, adversely affecting interfacial heat transport. The temperature jump $\Delta T$ can be directly determined from the temperature profile, and the heat flux $J_q$ can be calculated by extracting the cumulative energy added/removed from the thermostat. With knowledge of both the heat flux and temperature jump, the interfacial thermal resistance (ITR) can be directly calculated using Eq. (\ref{eq:Rk}).
\begin{figure}[h!]
\centering
\includegraphics[width=0.85\textwidth]{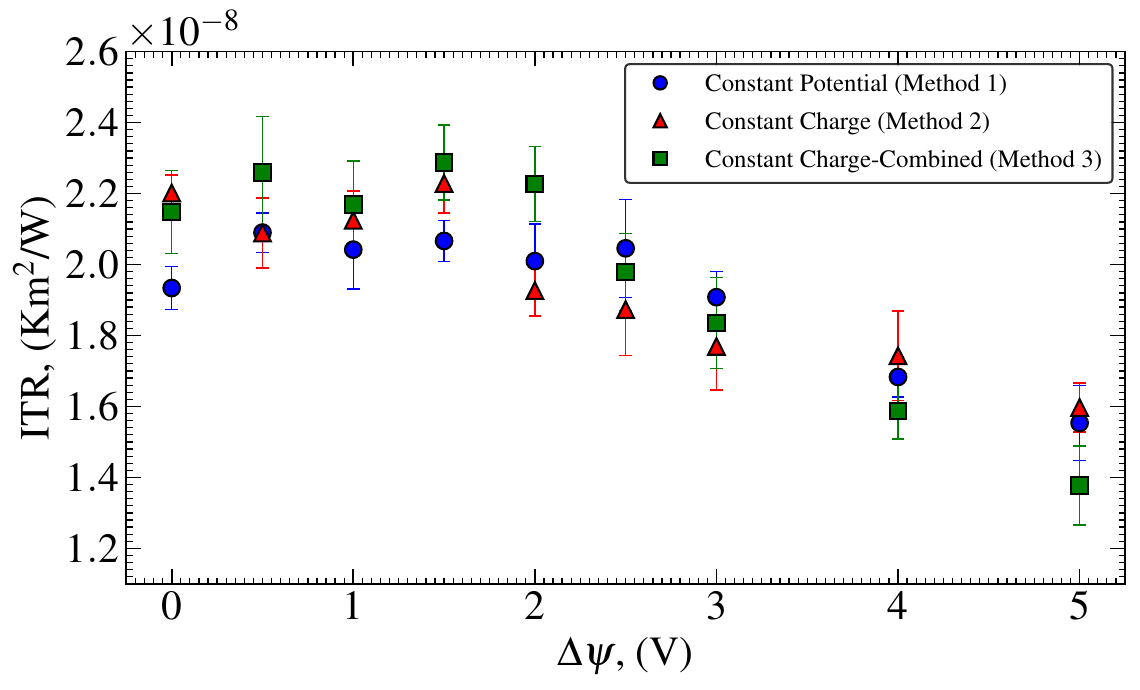}
\caption{Comparison of the ITR as a function of the potential difference for the graphene supercapacitor calculated using the three methods (Method 1 - constant potential, Method 2 - constant charge, and Method 3 - constant charge (combined)) . The error bars are the standard errors in the ITR from five independent simulations with different initial configurations. }
\label{ITR}
\end{figure}

Figure \ref{ITR} compares the ITR as a function of potential difference, calculated using the three methods described above. They are found to have overlapping error bars (shown as one standard error in the mean of five independent simulations), and no systematic differences can be seen between any of the modelling methods used. All three methods produced similar qualitative trends of the ITR being constant between 0 and 2 V and decreasing with an increasing potential difference after that.

\par
Furthermore, as shown in Fig. \ref{ITR}, the ITR appears to be relatively insensitive to potential differences at lower values and decreases as the potential difference increases. In a computational study by Alexeev \textit{et al.} \cite{alexeev2015kapitza}, it was reported that the ITR strongly depends on the height of the first water density peak at a graphene-water interface. However, for the graphene-IL interface, the first density peak height alone does not fully explain the trend of ITR variation with increasing potential difference. Alosious \textit{et al.} \cite{alosious2021nanoconfinement} later reported that the area density factor (the ratio of the number of molecules in a fluid slab adjacent to the interface, to the interface's surface area) is a primary factor in determining the ITR at a CNT-water interface as it accounts for curvature. Similarly,  we consider the total mass in a slab of IL adjacent to the graphene electrode (for planar interfaces, the surface area will be the same in all cases). The slab's thickness is chosen to accommodate the first few density peaks, unlike Alexeev \textit{et al.} \cite{alexeev2015kapitza} who consider only the first density peak. Analysis of the normalized slab mass revealed that the trend in slab mass variation with an increasing potential difference is inversely related to the ITR's dependence on potential difference (see Supporting Information for slab mass analysis). We propose that this is because a higher slab mass enhances heat transfer at the interface, reducing the ITR. This result suggests that the interfacial thermal transport in graphene-IL interfaces is not just a function of the first density peak but is also influenced by the successive peaks, especially for higher electrode charge density. As the potential difference increases, the surface charge density on the electrode surface increases, resulting in additional Coulombic interactions between the electrode and electrolyte in addition to van der Waals interactions. The influence of this electrostatic interaction is not only limited to the first layer but also to the successive layers of IL. Furthermore, it has been observed that the ITR exhibits variations between the anode and cathode interfaces, which can be attributed to the asymmetrical ion structure near the interface when a potential difference is introduced.
\begin{figure}[h!]
\centering
\includegraphics[width=0.8\textwidth]{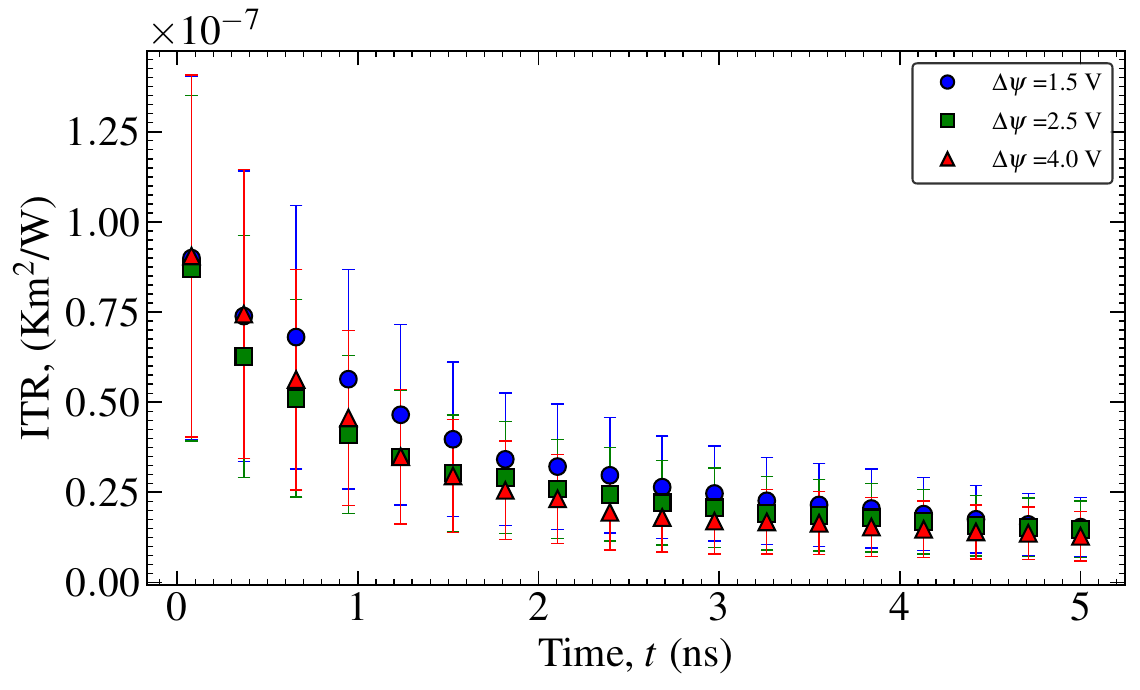}
\caption{The variation in ITR during the charging process of the graphene supercapacitor using CPM (Method 1). The error bars are the standard error in the mean of five independent simulations. }
\label{charging}
\end{figure}
\par
Although the CCM has been deemed adequate for studying interfacial thermal transport in supercapacitors, a CPM is necessary to comprehend the variations in ITR during the charging and discharging process. The utilization of CPM allows us to observe the changes in ITR during the charging process, as shown in Fig. \ref{charging}. Since the surface charge density of the electrode influences the interfacial thermal transport, the ITR is expected to decrease with an increase in potential difference. Ideally, the ITR is calculated using the time-averaged (entire production run) heat flux and temperature jump; however, only short intervals are available for computing averaged value during the charging process, leading to large statistical uncertainties in the ITR. Consequently, a qualitative inference is favored over quantitative results in this particular case. It is critical to consider that the interfacial thermal resistance and heat transfer vary throughout the charging and discharging process when designing and modeling supercapacitor interfaces and estimating heat dissipation.
\subsection{Graphene/CNT Composite Electrodes}

\begin{figure}[b!]
\centering
\includegraphics[width=1\textwidth]{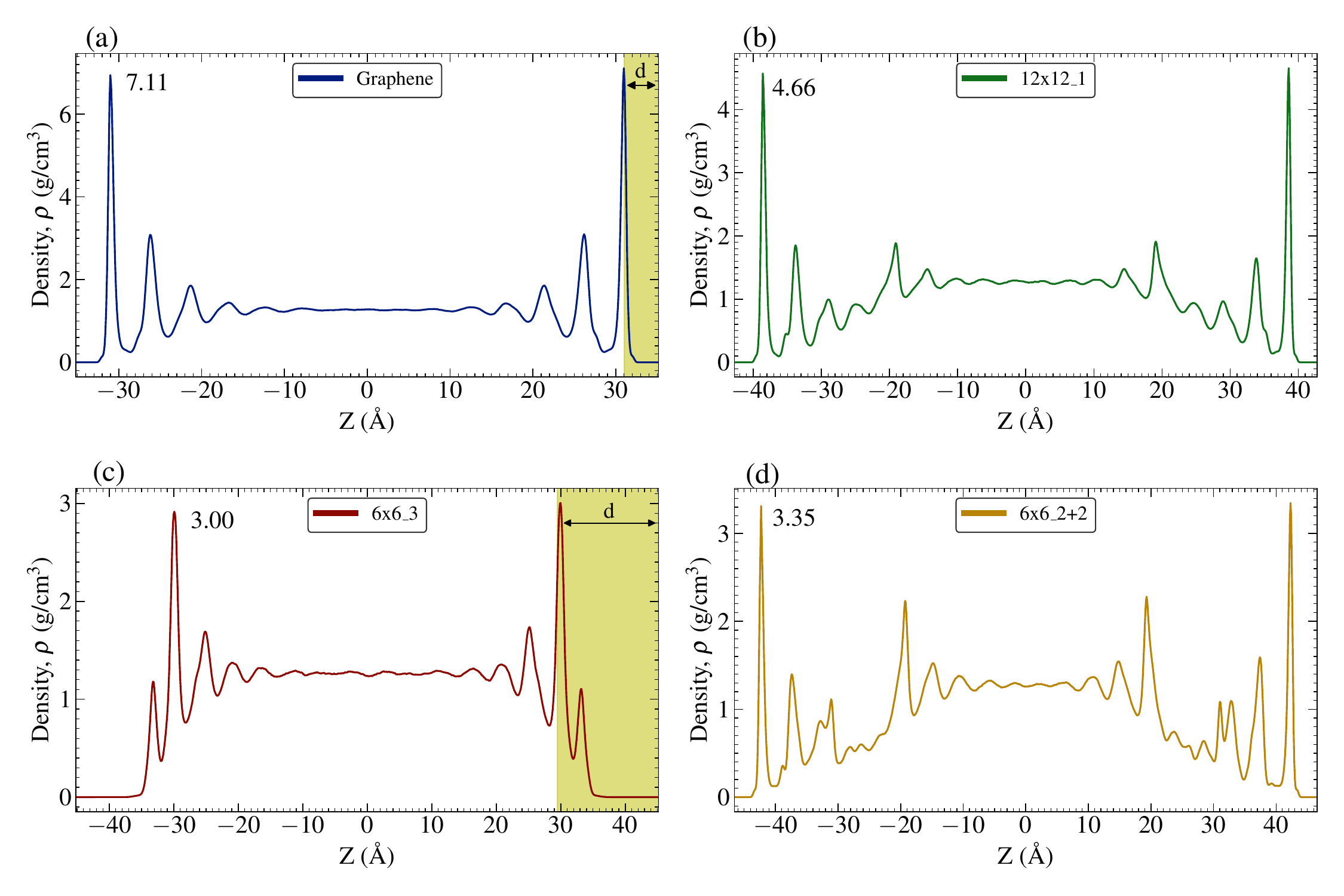}
\caption{The density profiles of the IL liquid for a few selected systems. (a) Planar graphene. (b) 12x12\_1 model. (c) 6x6\_3 model. (d) 6x6\_2+2 model. The height of the first liquid density peak is marked near the peak. The distance from the graphene layer to the first density peak is denoted as "d". }
\label{density2}
\end{figure}
The third part of this study focuses on examining a supercapacitor's electrical and thermal characteristics by employing diverse combinations of graphene/CNT composite electrodes. Molecular dynamics simulations were carried out using Method 1, as discussed in the methodology section. Figure \ref{density2} displays the density profiles of IL for a select set of systems. A typical density profile of IL confined between uncharged planar graphene sheets is shown in Fig. \ref{density2}a. The height of the first density peak is about 6.5 g cm$^{-3}$, and the interfacial layering can be seen up to 25 \AA\ from the graphene surface. When a single CNT of diameter 16.5 \AA\ (12,12) is attached horizontally to the graphene, the structure of IL near the electrode is changed and is shown in Fig. \ref{density2}b. The first density peak is reduced to 5 g cm$^{-3}$ due to the reduced contact between IL and graphene in the presence of CNT. Moreover, an additional density peak adjacent to the CNT emerges, and the interfacial layering region extends to 40 \AA. Figure \ref{density2}c illustrates the IL structure when three (6,6) CNTs are stacked horizontally on the graphene. The density peak is shifted towards the bulk portion,  indicating that the contact between IL and graphene is shadowed by the CNTs.  The distance between graphene and the first density peak, $d$ increased from 4 \AA\ to 16 \AA\ compared to the planar graphene electrode. In addition to the diminished contact between graphene and IL, the height of the density peak is also reduced compared to planar graphene, which shows poor solid-fluid wetting despite having additional surface area. Similarly, Fig. \ref{density2}d shows the structure of IL for a two-layer system. Here also, we can see a reduction in the maximum density peak and large layering region due to the filling of IL inside the porous structure. However, compared to planar graphene, a higher number of density peaks can be seen, suggesting an enhancement in interfacial wetting. Similar structural patterns can also be observed in vertical arrangements and other horizontal systems. 
\begin{figure}[b!]
\centering
\includegraphics[width=1\textwidth]{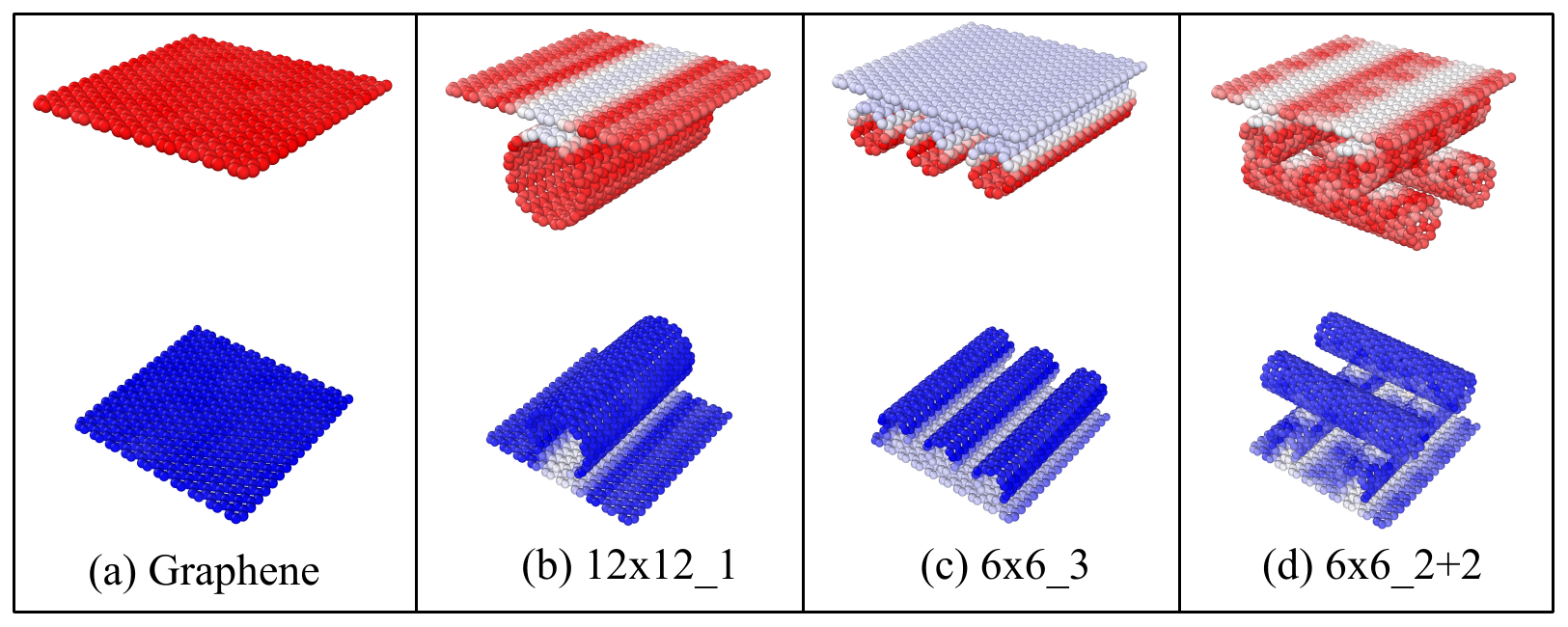}
\caption{The surface charge distribution on the electrode induced by
the cations and anions averaged over the entire simulation time with the CPM method. (a) Planar graphene. (b) 12x12\_1 model. (c) 6x6\_3 model. (d) 6x6\_2+2 model.}
\label{contour}
\end{figure}
\par
From the above discussion, it is clear that the structure of the IL and the contact between the electrode and electrolyte varies depending on the morphology of the electrode. This could lead to uneven charge distribution on the electrode surface. Therefore, a surface charge contour of the electrodes is analyzed to understand the distribution of induced charge on the electrodes. Figure \ref{contour} shows the surface charge distribution on the electrode induced by the cations and anions averaged over the entire simulation time (using CPM). The surface charge distribution of all the electrodes can be found in the Supporting Information. The induced charge on each atom is measured and averaged over the whole simulation time used to plot the charge contour. Blue and red colors represent positive and negative charges, respectively, while white indicates zero charge. For a planar graphene electrode, the time average surface charge density is equally distributed over the entire graphene surface (without any white spots), as illustrated in Fig. \ref{contour}a. This shows a very efficient contact between the electrode and electrolyte. However, for a graphene/CNT composite electrode, some areas with almost zero charge density (depicted as white spots in Fig. \ref{contour}b-c) suggest that there is a portion of the electrode (contact point of graphene and CNT) that is inaccessible to the IL. For instance, a large area with zero charge is visible for the three graphene/CNT electrode system (Fig. \ref{contour}c), indicating an inefficient contact between the electrode and electrolyte. Since the electrode is periodic in \textit{x} and \textit{y} directions, the number of CNTs effectively means the distance between the CNTs. Here the entire graphene surface and about half of the CNT surface are inaccessible to the IL, resulting in an inefficient electrode-electrolyte contact. Similarly,  Fig.  \ref{contour}d shows the charge contour for a two-layer CNT electrode system in which the white regions are visible on different electrode positions. This particular stacking arrangement mimics a porous electrode structure, with IL filling the pores. The charge is induced on the electrode structure except for the contact points between graphene and CNTs. The surface charge density contours in Fig. \ref{contour} provide a qualitative understanding of the interaction between the electrolyte and various electrode structures.
\par
The addition of CNTs in different dimensions and arrangements will increase the total surface area of the electrode. However, it should be noted that an increase in the total surface area does not necessarily equate to a proportional increase in the area that is accessible to the electrolyte, as this is dependent on the electrode morphology. To assess the efficacy of each electrode, the accessible surface area (ASA) is calculated for all of the electrodes. ASA is calculated using the "rolling ball" algorithm, which probes the surface and pores of the electrode with a sphere equivalent to the size of ions \cite{shrake1973environment}. We selected a radius of 2.5 \AA\, which is comparable to the size of the PF$_6^-$ anion. To establish whether a point is accessible or buried, all points are examined on the surfaces of nearby atoms. Finally, the ASA is calculated by multiplying all the accessible points by the surface area of each point. Also, the total surface area (TSA) is calculated by adding the surface area of graphene and the lateral surface area of CNTs.

\begin{figure}[t!]
\centering
\includegraphics[width=1\textwidth]{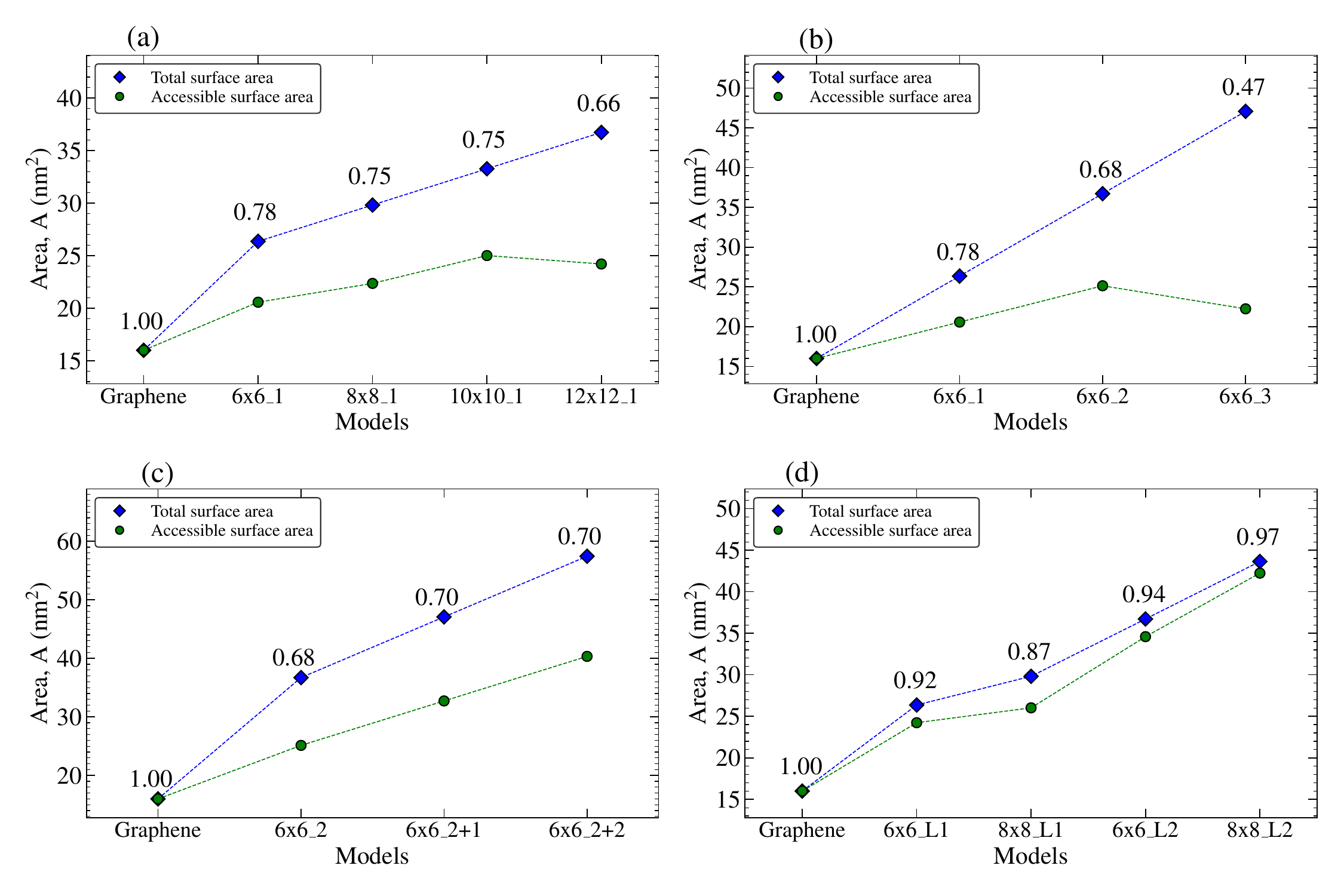}
\caption{Comparison of accessible surface area (ASA) and total surface area (TSA) for all the electrodes classified based on (a) CNT diameter, (b) number of CNTs, (c) the number of CNT layers, and (d) vertical arrangement of CNTs. The ratio between ASA and TSA for all the electrodes is also denoted. }
\label{area}
\end{figure}

Figure \ref{area} compares the TSA with ASA for all the electrodes, and their ratio is also indicated. The TSA is always higher than the ASA, as it represents the maximum available surface area. From here on, the comparison of different properties is classified into four categories based on diameter, number, layer, and alignment of CNTs on graphene. An increase in the diameter of CNT results in an increase in both TSA and ASA, and the ratio is about 0.79 for all four diameters, as depicted in Fig. \ref{area}a. From Fig. \ref{area}b, it is clear that adding two (6,6) CNTs in parallel increases the ASA, but adding one more CNT drastically reduces the ASA to 0.55, indicating that almost half of the electrode surface is inaccessible to the electrolyte. Therefore, increasing the number of CNTs or closely packing them will limit filling of the pores by the electrolyte. On the other hand, Fig. \ref{area}c shows that stacking more layers increases the ASA without compromising the ASA/TSA ratio. Stacking CNTs one above another can control the porosity of the electrode, thereby optimizing the electrode structure as per requirements. Finally, the ASA for the vertical arrangement is illustrated in Fig. \ref{area}d. Since the CNT is vertically positioned, the inside volume of CNT is also accessible to the ions if the ion size is smaller than the CNT diameter. Also, there are no bulky contact points between graphene and CNTs, unlike in the horizontal arrangement. Therefore, the ASA/TSA ratio is higher for vertical than horizontal stacking. 
\subsubsection{Selective ion entry}

\begin{figure}[b!]
\centering
\includegraphics[width=0.9\textwidth]{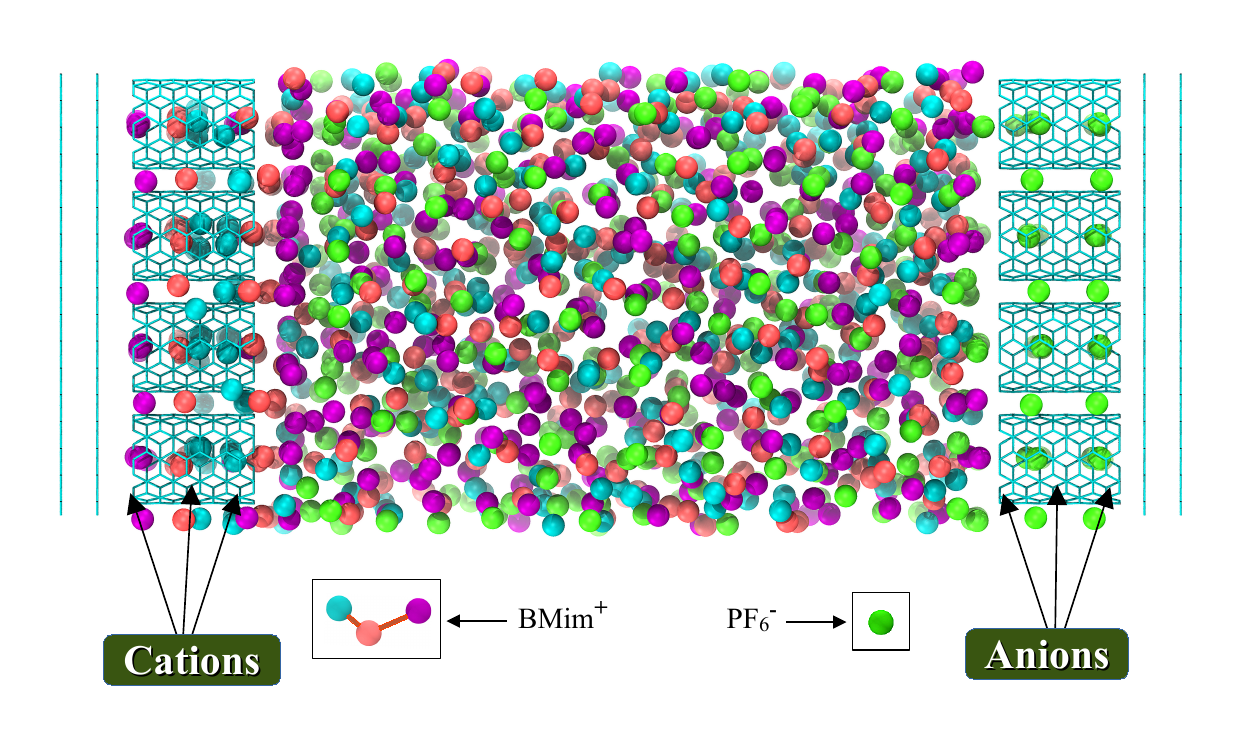}
\caption{Schematic diagram of the closely packed vertical CNT electrode system. The selective entry of cations inside anode pores and anodes inside cathode pores under 4.0 V potential difference is depicted. }
\label{selective}
\end{figure}
When a carbon nanotube of sufficiently high diameter is oriented perpendicular to graphene, ions can penetrate inside it. It was found that for the (8,8) CNT, both anion and cation can enter the CNT at zero potential difference. However, for (6,6) CNT,  the diameter is not big enough for the ions to enter inside at zero potential. Nevertheless, when a potential difference is applied across the supercapacitor, we found that cations and anions respectively enter the CNTs connected to the anode and cathode. To better understand the dependence of CNT filling on potential difference, another electrode was modeled with 16 (6,6) CNTs vertically aligned, as illustrated in Fig. \ref{selective}. This closely packed arrangement ensures the ions cannot pass through the gap between CNTs. This system is also equilibrated under 1 bar pressure, similar to other models. When the potential difference is zero, none of the ions can enter the CNTs or the gap between CNTs. However, when a potential difference is applied, all the anode CNTs are filled with cations and vice versa. Fig. \ref{selective}  shows that only the anions are inside the cathode CNTs and cations inside the anode CNTs. This way, we can completely separate the anions and cations and attach them to the cathode and anode. This mechanism can be achieved by choosing the correct CNT diameter considering the size of the ions. Density profiles of both anions and cations, with and without applied potential, are provided separately in the Supporting Information. Even though a perfect double-layer (ideal case for an EDLC supercapacitor) can be achieved on both electrodes, the specific capacitance of the supercapacitor cannot be improved due to ineffective contact between the electrode and electrolyte. Since the CNTs are tightly packed, the accessible surface area for ions is much less, making this type of arrangement ineffective in improving the performance of a supercapacitor. Even though this electrode morphology is unsuitable for supercapacitors, this technique can be used in other applications, such as desalination and ion separation \cite{saffarimiandoab2022molecular}. While this study utilizes a coarse-grained model of the IL instead of an atomistic one, it is expected that the same phenomenon will still be observed, despite a potentially different quantitative trend for the physical system.

\subsubsection{Electrical properties}

One of the primary objectives of this work is to compare the electrical properties, such as capacitance and energy density of the supercapacitor, by using different electrode morphologies. The double layer capacitance of a supercapacitor primarily depends on the potential drop ($\Delta \psi$   = $\psi^{cathode}-\psi^{anode}$) across the cell and the surface charge density ($\sigma$) of the electrode. The potential drop is considered as the diﬀerence between the electrostatic potential drop in the charged capacitors and discharged capacitors,
\begin{equation}
 \Delta \Delta \psi   = \Delta \psi^{charge}-\Delta \psi^{uncharge}.
\end{equation}
Whereas the surface charge density is the average surface charge on the electrode divided by the surface area of the electrode. The surface can be either the total surface area (TSA) or the accessible surface area (ASA), and the following analyses consider ASA for all the calculations unless specified. Usually, capacitance can be classified into two categories: integral and differential capacitance. Integral capacitance is calculated by taking the ratio of average charge density and the potential drop across the capacitor,
\begin{equation}
C_{I}=\frac{\vert\sigma\vert}{\Delta \Delta \psi}.
\end{equation}
The differential capacitance from the rate of change of charge density as a function of potential drop,
\begin{equation}
C_{D}=\frac{d \sigma}{ d\psi}.
\end{equation}

\begin{figure}[t!]
\centering
\includegraphics[width=1\textwidth]{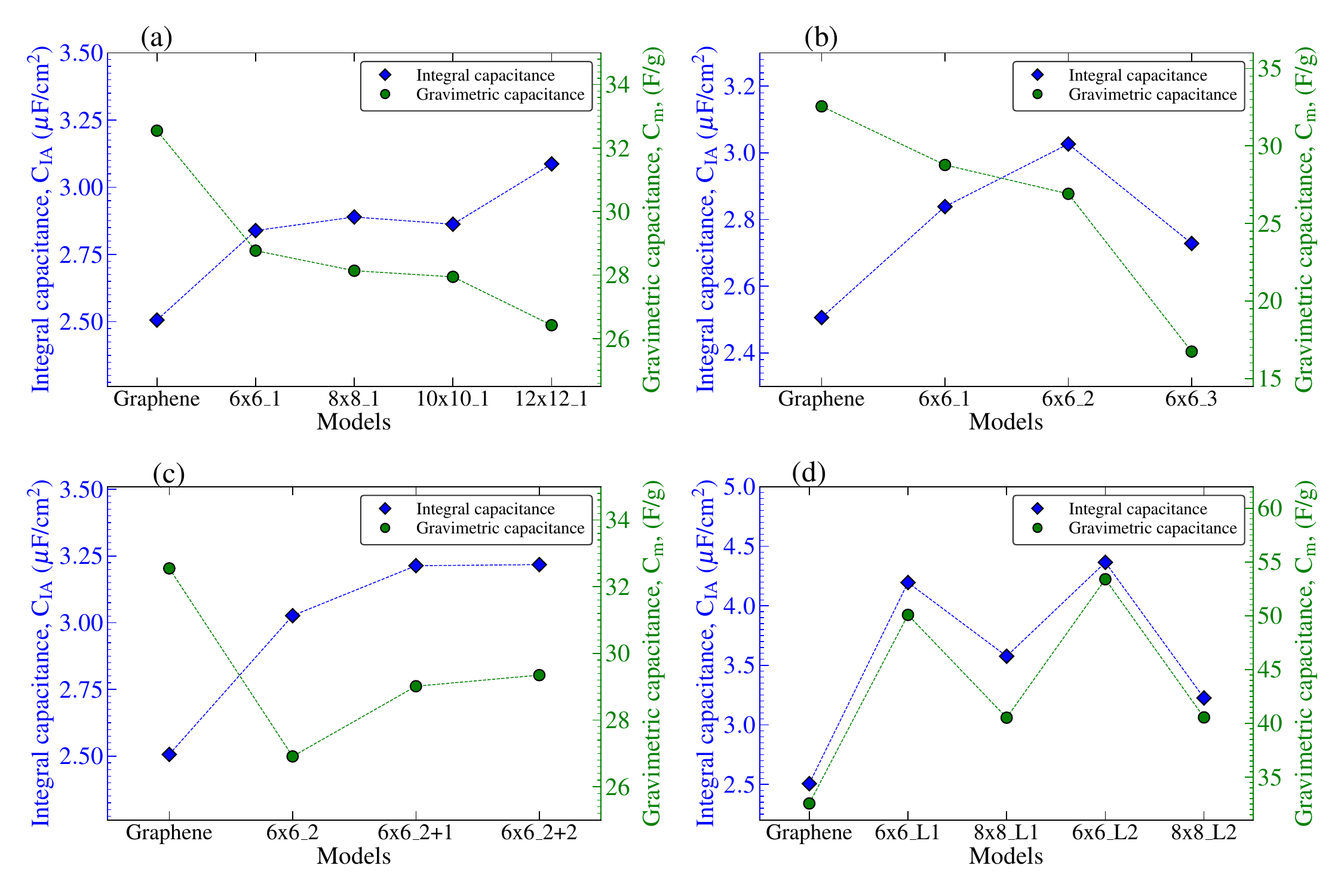}
\caption{Integral capacitance, C$_{IA}$ (Calculated based on ASA) and Gravimetric capacitance, C$_m$ (calculated based on the TSA) for all the electrode models classified based on (a) CNT diameter, (b) number of CNTs, (c) the number of CNT layers, and (d) vertical arrangement of CNTs.}
\label{mass}
\end{figure}
Since this study focuses on the overall electrical performance of the supercapacitor for different electrode structures, integral capacitance is used throughout this paper. Figure \ref{mass} shows the comparison of integral capacitance and gravimetric capacitance for different electrode structures, where the gravimetric capacitance is calculated from the total mass of carbon atoms making up each electrode. Adding horizontal CNTs lowers the gravimetric capacitance compared to planar graphene, indicating reduced contact efficiency between the electrode and electrolyte. The integral capacitance still increases with the addition of one- or two-CNT horizontal layers, showing that overall performance still improves, but the three-CNT horizontal layer reduces both the integral and gravimetric capacitance, indicating that ions are obstructed by too many layers of closely-packed CNTs. From Fig. \ref{mass}c, we can see that adding a second layer of CNT will modestly increase gravimetric and integral capacitance compared to a single layer of horizontal CNTs. The capacitance with vertically-added CNTs is shown in Fig. \ref{mass}d, which shows overall higher gravimetric and integral capacitance compared to planar graphene. Indeed, the vertical arrangement of 6x6 CNTs of length 2nm (6x6\_2L) has the highest capacitance value among all the electrodes studied.
\par
\begin{figure}[t!]
\centering
\includegraphics[width=1\textwidth]{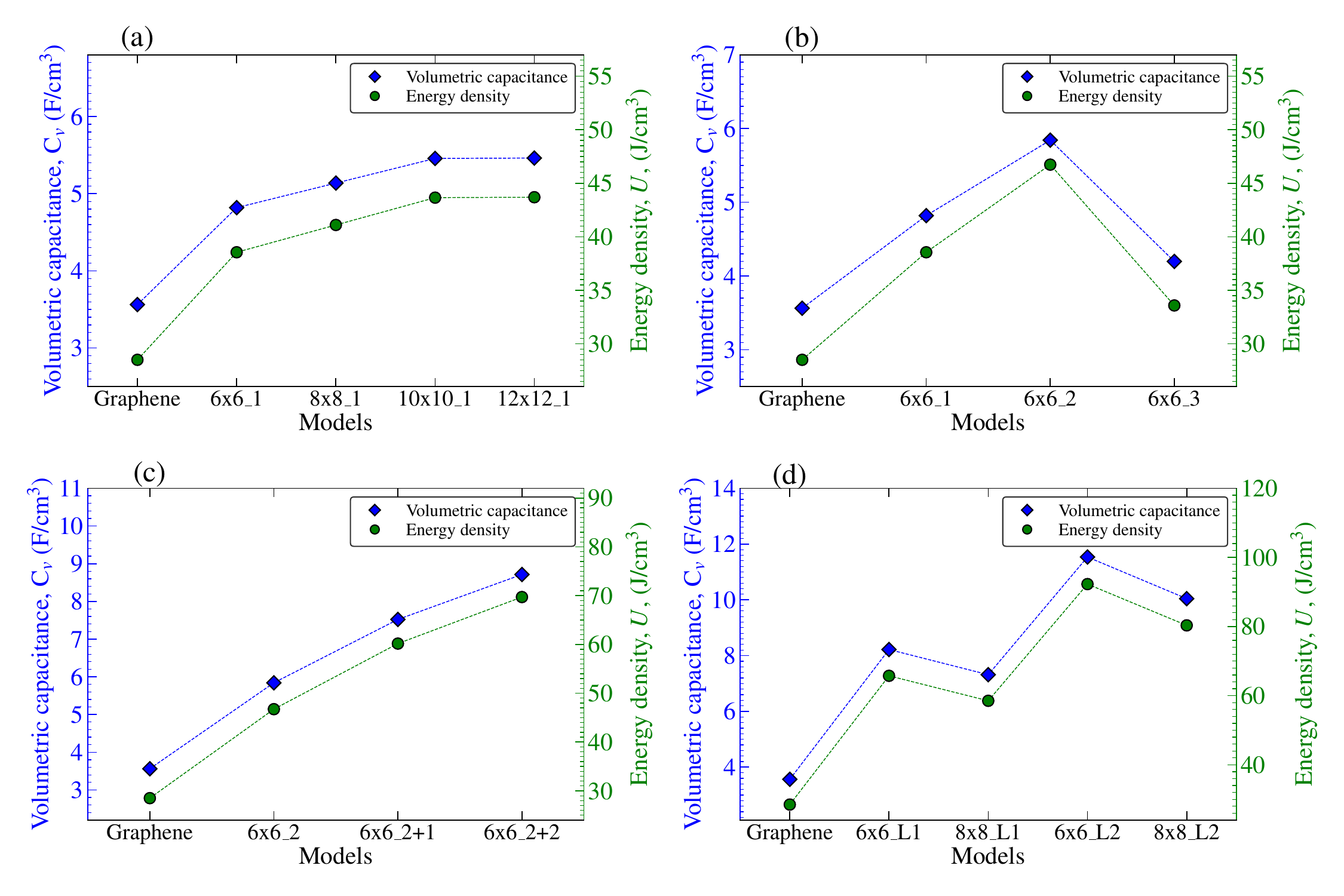}
\caption{Volumetric capacitance, C$_{v}$ and Energy density, $U$ for all the electrode models classified based on (a) CNT diameter, (b) number of CNTs, (c) the number of CNT layers, and (d) vertical arrangement of CNTs.}
\label{energy}
\end{figure}
Furthermore, we have computed the volumetric capacitance and energy density of the supercapacitor for all electrodes and illustrated them in Fig. \ref{energy}. The volumetric capacitance is given by 

\begin{equation}
C_v=\frac{C}{V}
\end{equation}
The energy density is given by
\begin{equation}
U=\frac{C_I.[\Delta \Delta \psi]^2}{2V},
\end{equation}
where $V$ is the volume of the supercapacitor.  Here the volume is calculated by considering the box size of each supercapacitor model. From Fig. \ref{energy}a, we can see that the volumetric capacitance and energy density increases with an increase in the diameter and converges above 10x10 CNT. This indicates that increasing the diameter after a particular value gives no advantage. When it comes to an increase in the number of CNTs, adding a third CNT or decreasing the distance between CNTs drastically lowers the volumetric capacitance and energy density, as depicted in Fig. \ref{energy}b. The electrode with a second CNT layer, however, enhances the volumetric capacitance and energy density.  Additionally, for vertical systems, smaller CNT diameters and longer lengths lead to better performance. These findings suggest that vertical CNT stacking is superior to horizontal stacking since it provides the ions with better access to almost the entire CNT surface, resulting in improved electrical properties.  However, the disadvantage of the vertical system is the difficulty in manufacturing such electrodes and scalability. In contrast, for horizontal systems, the most effective electrode is the two-layer configuration, which is relatively easy to produce and can be scaled up by increasing the number of layers.

\subsubsection{Thermal properties}

\begin{figure}[b!]
\centering
\includegraphics[width=1\textwidth]{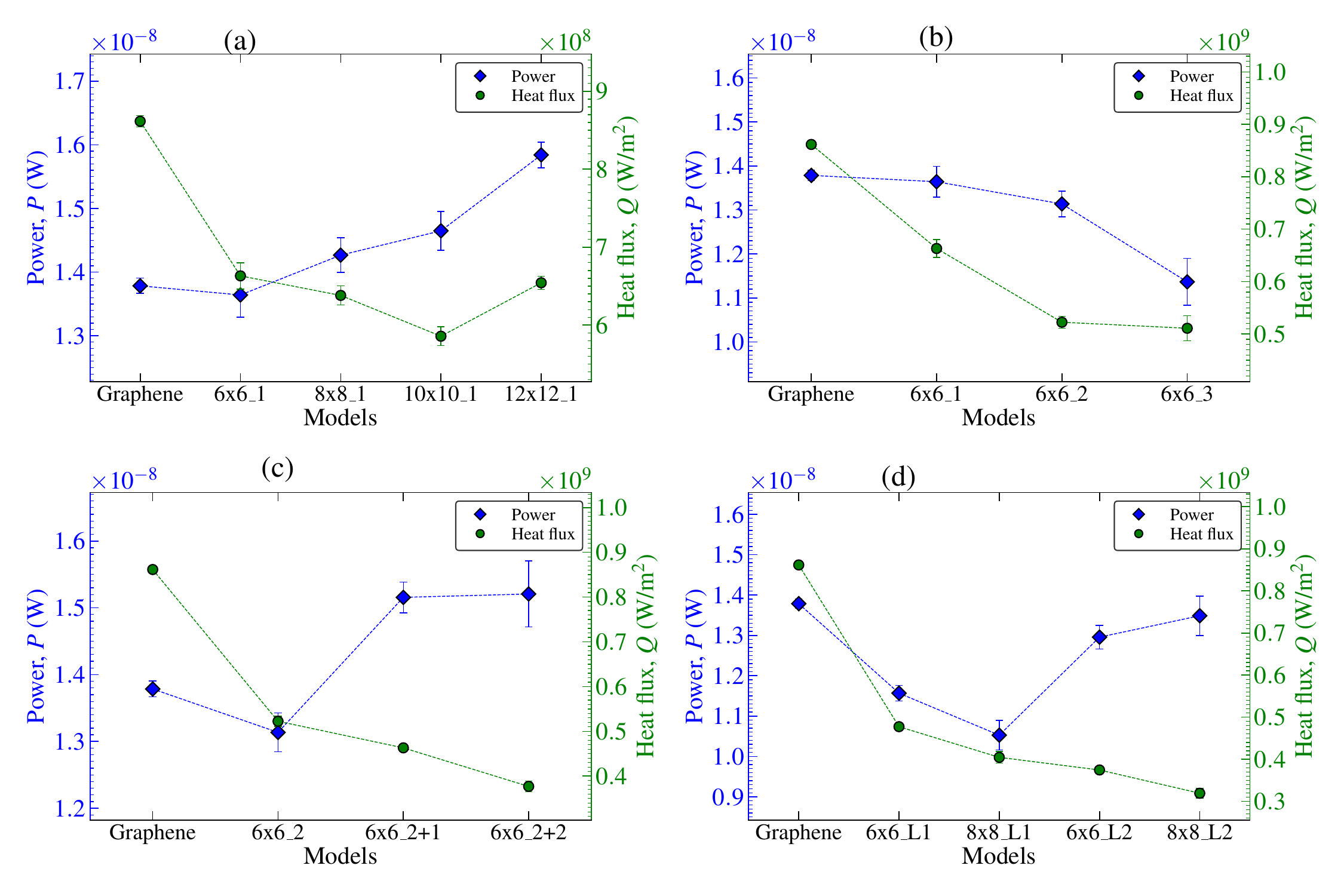}
\caption{Thermal power output, $P$ and heat flux $Q$ for all the electrode models classified based on (a) CNT diameter, (b) number of CNTs, (c) the number of CNT layers, and (d) vertical arrangement of CNTs.}
\label{heatflux}
\end{figure}

To ensure maximum performance of the supercapacitor, it is crucial to maintain its temperature within a specified range. Efficient dissipation of heat generated within the supercapacitor is necessary to maintain the temperature. Hence, interfacial thermal transport, which plays a significant role in heat dissipation, must be considered while designing the electrode. The interfacial thermal resistance or Kapitza resistance is the primary parameter determining the heat transfer property of an interface \cite{alosious2019prediction}. However, in this study, the interface between the electrode and electrolyte cannot be defined due to the complexity of the electrode structure. Therefore, the heat transfer performance of each electrode is calculated based on the power and the heat flux dissipated from each electrode. The power output is calculated by measuring the average cumulative energy added and removed from the thermostat divided by the simulation time. On the other hand, the heat flux is calculated by dividing the heat power output by the accessible surface area of the electrode.

The heat flux and power output for all the electrodes are illustrated in Fig. \ref{heatflux}. It is observed that the planar graphene electrode has the highest heat flux of all the electrodes. Here the heat flux shows how effectively the heat transfer takes place at the electrode. A higher heat flux value for graphene, and lower heat flux values when adding CNTs, indicates that adding CNTs will increase the interfacial resistance to heat flow, as a result of worse contact between CNTs and electrolyte. However, the heat power output trend shows that, despite increasing the interfacial thermal resistance, the total thermal energy dissipated still increases when CNTs are added due to the overall increase in surface area. As depicted in Fig. \ref{heatflux}a, power output increases with the diameter of the CNT, while heat flux decreases until 10x10, then slightly increases with further diameter increase. From Fig. \ref{heatflux}b, it is clear that when the number of CNTs increases, the heat flux and power output decrease, indicating an unfavorable design for thermal aspects. This reduction can be attributed to the improper contact between the electrode and electrolyte leads to inefficient thermal transport. Interestingly, adding a second layer of CNTs improves the heat transfer in addition to the better electrical performance as depicted in Fig. \ref{heatflux}c. Even though interfacial thermal resistance is reduced, adding CNTs increases the total surface available for heat transfer, resulting in a visible increase in power in most horizontal systems. Figure \ref{heatflux}d displays that power and heat flux are lower in vertical systems than in graphene, indicating that the heat transfer efficiency is not advantageous for the vertical arrangement of CNTs. Furthermore, longer CNTs exhibit superior power dissipation compared to shorter ones.
\par

To gain a deeper comprehension of the heat transfer mechanism at the electrode-electrolyte interface, we have also investigated the vibrational density of states (VDOS) of IL and all the electrodes.
The calculation and results of the VDOS is described in the Supporting Information. The VDOS overlap between electrodes showed no significant change, suggesting that the vibrational coupling between the electrode and IL remains unaffected by changes in electrode structure. Hence, the observed alteration in interfacial thermal transport due to the inclusion of CNTs cannot be attributed to changes in the vibrational spectrum, as noted in earlier studies \cite{alexeev2015kapitza,alosious2022effects}. 
Consequently, the change in interfacial thermal transport can be attributed to the electrolyte structure adjacent to the electrode interface. As discussed earlier, the height of the density peak or the mass of electrolyte present near the electrode can be used to understand the variation in heat transfer with each electrode. From Fig. \ref{density2}, it is clear that the height of the first density peak is highest for graphene, indicating a higher wetting coefficient leading to lower thermal resistance. Therefore, the interfacial thermal transport between the electrode and electrolyte can be predicted by knowing how density packed is the ions in and around the pores of the electrode. 
\subsubsection{Optimum electrode structure}

\begin{figure}[b!]
\centering
\includegraphics[width=.9\textwidth]{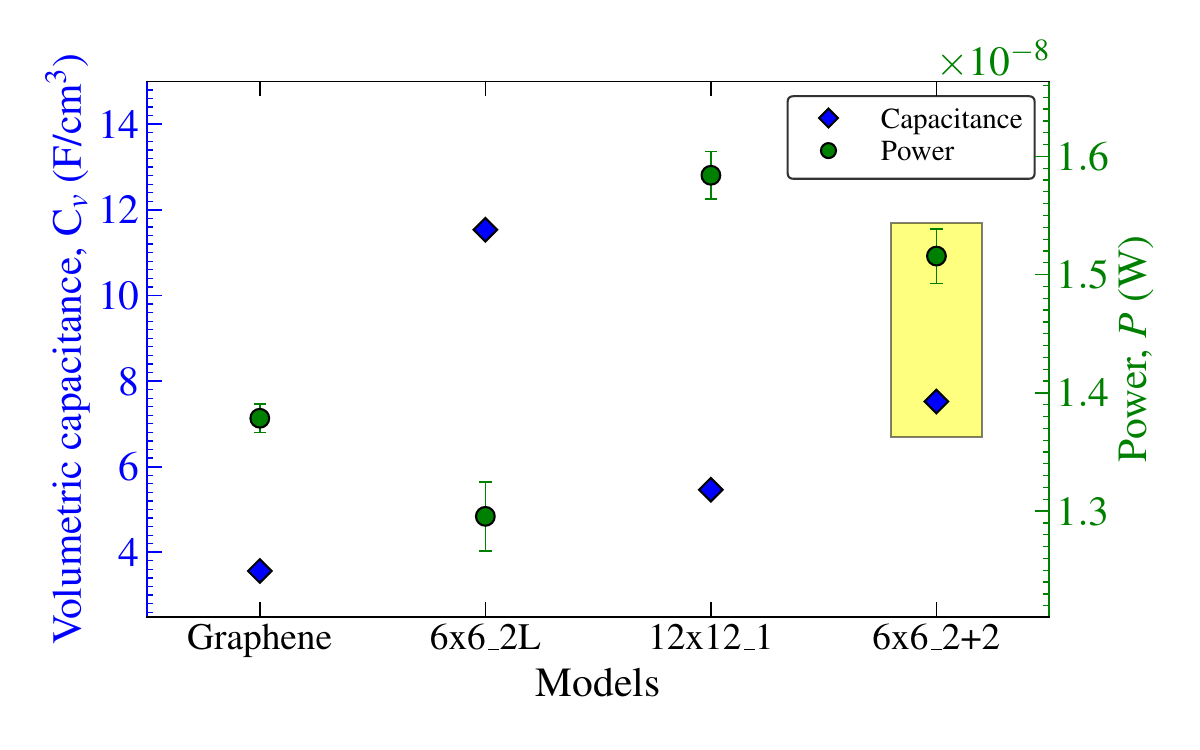}
\caption{Volumetric capacitance, C$_{v}$ and thermal power output, $P$, compared for selected electrodes to quantify electrical and thermal performance, with metrics included for graphene as a comparison point. }
\label{optimum}
\end{figure}

Finally, we try to identify an electrode structure that maximizes both electrical and thermal performance, based on volumetric capacitance and heat dissipation respectively (Fig. \ref{optimum}). Of all the electrode structures considered, the vertical arrangement of 6x6 CNTs with 2 nm length (6x6\_2L) has the highest volumetric capacitance of approximately 13.6 F cm$^{-3}$, but its thermal performance is inferior to all other electrodes. Similarly, the single CNT system with a 12x12 chirality (12x12\_1) has the highest power output of approximately 1.58 $\times$ 10$^{-8}$ W, but its electrical performance is the worst among the chosen electrodes. Therefore, it is clear that the electrode structure with the highest electrical performance has poor thermal performance and vice versa. However, if we look at the two-layer system (6x6\_2+2), the volumetric capacitance and power output values are the second highest in the selected electrodes. The volumetric capacitance of the two-layer system is 7.5 F cm$^{-3}$, and the power output is 1.52 $\times$ 10$^{-8}$ W. Compared to the 12x12\_1 electrode with the highest power output, only about 3.8\% lesser in power output is visible for the two-layer system. Therefore, the 12x12\_1 system can be eliminated and replaced with the two-layer system. Nonetheless, compared to the vertical system (6x6\_2L) with the highest capacitance, the two-layer system has a 44.8\% lesser capacitance value. Therefore, we cannot completely eliminate the vertical system regardless of its poor thermal performance. Vertically stacked CNT structures can be used if electrical performance is the only requirement. However, before concluding, other factors, such as the structural balance, production, and scalability of the electrode structure, must also be taken into account.
\par
Manufacturing vertically aligned electrodes is challenging due to limited contact points with the graphene and difficulty in maintaining the vertical alignment of CNTs. Structural stability is also a concern, and scaling up the thickness of the electrode is restricted by the length of CNT. However, these issues do not apply to a two-layer electrode system that uses horizontal stacking of CNTs. This system offers higher structural stability, easier manufacturing, and the ability to adjust the pore size by controlling the distance between CNTs. Considering these factors, the two-layer (multi-layer) electrode system is the optimal structure with better electrical and thermal performance. Various techniques such as spin-casting \cite{shakir2014high}, self-assembly/solution phase processing \cite{qiu2010dispersing,cheng2011graphene}, and chemical vapor deposition (CVD) methods \cite{fan2010three,niu2020enhanced} can be used to manufacture graphene/CNT composite electrodes. The multi-layer structure can be obtained by functionalizing the surface of graphene and CNT with opposite charge groups, which enables electrostatic attraction between CNT and graphene \cite{yu2010self,byon2011thin}.

\section{CONCLUSIONS}

In summary, we have investigated the interfacial thermal transport in a graphene-ionic liquid supercapacitor, comparing constant charge and constant potential method molecular dynamics simulations. To carry out the new constant potential simulations while enabling thermal transport, we use a novel, computationally efficient approach where the charge sites are on lattice points but the atoms are allowed to move.  Our results indicated that the constant potential and constant charge methods produce similar results for planar graphene electrodes, because the IL structure is consistent between both methods.  Furthermore, the interfacial thermal resistance decreases as the electrode potential difference increases due to the increased density of ions close to the interface. Even though both methods give similar results in planar electrodes, the constant charge method is unsuitable for porous electrodes, and only the constant potential method can be used to understand the heat transfer mechanism during the charging/discharging process.
\par
This study also aimed to identify the optimum electrode structure by modeling graphene/CNT composite electrodes and studying their electrical and thermal properties while varying the diameter, number, layers, and alignment of CNTs. It was observed that placing a CNT over the graphene layer led to a decrease in the accessible surface area for ions compared to planar graphene. Vertically aligned CNT electrodes exhibited superior electrical performance, such as specific capacitance and energy density, while their thermal performance was inferior to that of horizontally aligned CNTs. Likewise, horizontally aligned CNTs with 12x12 chirality and two-layer arranged electrode structures showed higher thermal performance. However, the single CNT with 12x12 chirality was not recommended due to poor electrical performance. Considering thermal and electrical aspects, the two-layer (multi-layer) system is the optimal electrode structure. For the two-layer model (6x6\_2+2), the volumetric capacitance increased by 108\%, and the thermal power output increased by 10\% compared to the planar graphene electrode. In future work, we will model different ionic liquids under various concentrations using all-atom models to understand the nanoscale mechanisms more deeply. The multi-layer electrode structure will also be further optimized with separately tuned pores for the cathode and anode, depending on the size of the anions and cations.

\section*{Acknowledgements}
The authors thank the Australian Research Council for its support for this project through the Discovery program (FL190100080). We would like to express our appreciation to Dr. Emily Kahl for her invaluable support in developing and debugging the source code used in this project. Additionally, we extend our thanks to the members of the Benhardt group for their feedback and suggestions, which greatly enriched our work. We acknowledge access to computational resources provided by the Pawsey Supercomputing Centre with funding from the Australian Government and the government of Western Australia, and the National Computational Infrastructure (NCI Australia), an NCRIS enabled capability supported by the Australian Government. We also acknowledge support and computer access through the Queensland Cyber Infrastructure Foundation (QCIF) and the Research Computing Centre at The University of Queensland.

\bibliography{reference}



\section*{Supplementary material}

# Supporting Information for "Interfacial Thermal Transport and Electrical Performances of Supercapacitors with Graphene/Carbon Nanotube Composite Electrodes"

Sobin Alosious,[†] Shern R. Tee,[‡] and Debra J. Searles[*,†,¶]

[†]Australian Institute of Bioengineering and Nanotechnology, The University of Queensland, Brisbane, QLD 4072, Australia

[‡]School of Environment and Science, Griffith University, Nathan, QLD, 4111, Australia

[¶]School of Chemistry and Molecular Biosciences, The University of Queensland, Brisbane, QLD 4072, Australia

E-mail: d.bernhardt@uq.edu.au

## Comparison of Interfacial Thermal Resistance for Different Systems

The normalized slab mass of the ionic liquid using a constant potential method (CPM - Method 1 in main text) is illustrated in Figure 1. The analysis of the slab mass demonstrates that the relationship between slab mass variation and increasing potential difference is inversely related to the dependence of the interfacial thermal resistance (ITR) on the po-

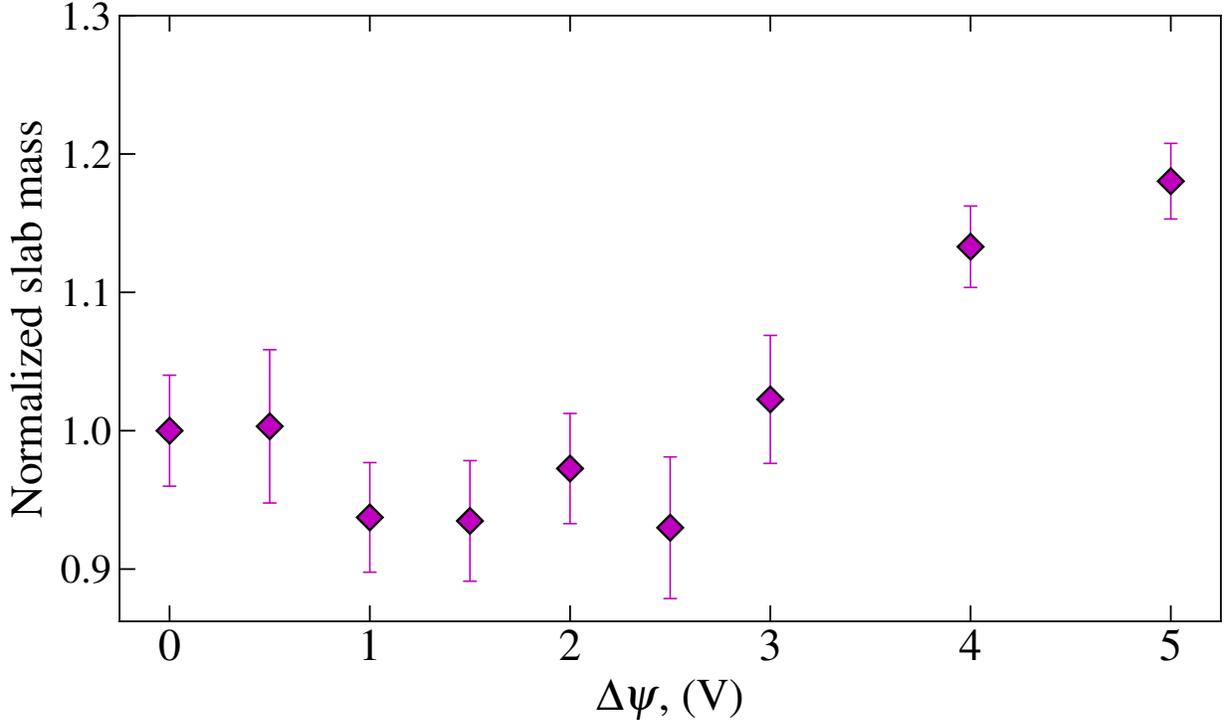

Figure 1: Normalized slab mass for different potential differences

tential difference. This is due to the fact that higher slab mass enhances heat transfer at the interface, thereby reducing the ITR. These findings indicate that the ITR at graphene-IL interfaces is not solely determined by the first density peak, but it is also influenced by the subsequent peaks, particularly for higher electrode charge density. With an increase in potential difference, the surface charge density on the electrode surface rises, leading to additional Coulombic interactions between the electrode and electrolyte in addition to van der Waals interactions. The impact of this electrostatic interaction extends beyond the first layer and affects the successive layers of IL.

## CNT/Graphene Composite Electrodes

Figure 2 displays the spatial distribution of surface charges on the electrode resulting from the presence of cations and anions, which have been averaged over the entire duration of the simulation using the CPM. The charge induced on each atom has been measured and

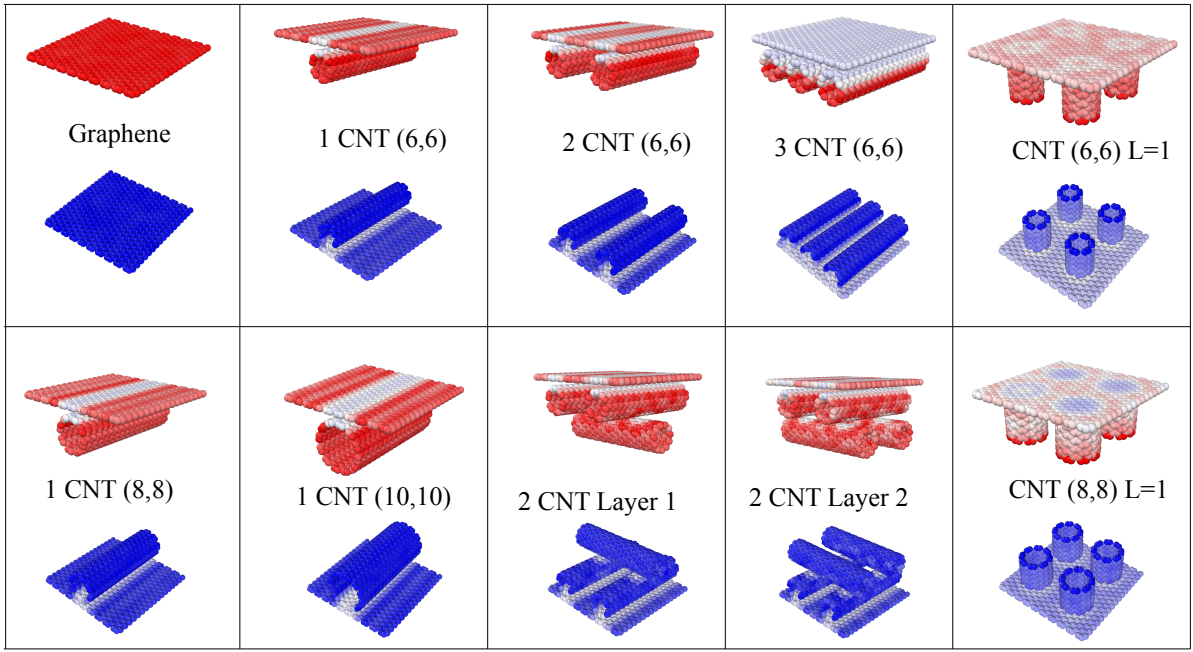

Figure 2: The surface charge distribution on all the electrodes induced by the cations and anions averaged over the entire simulation time with the CPM method.

averaged throughout the simulation time to generate the charge contour. In the contour plot, blue and red colors represent positive and negative charges, respectively, while white regions correspond to areas with zero charge. This visualization provides valuable insights into the electrostatic behavior of the system and highlights the distribution of charges on the electrode surface.

## Selective ion entry

Figure 3 illustrates the density profiles of anions and cations before and after the application of a potential difference. When the potential is zero, the carbon nanotube (CNT) pores are devoid of any ions, displaying zero density. However, upon the application of a potential, a pronounced density peak of anions becomes evident within the cathode CNT pores, whereas the anode pores remain unoccupied by anions. Similarly, Figure 4 presents the density of cations, which initially shows zero density within the pores. Nevertheless, upon the appli-

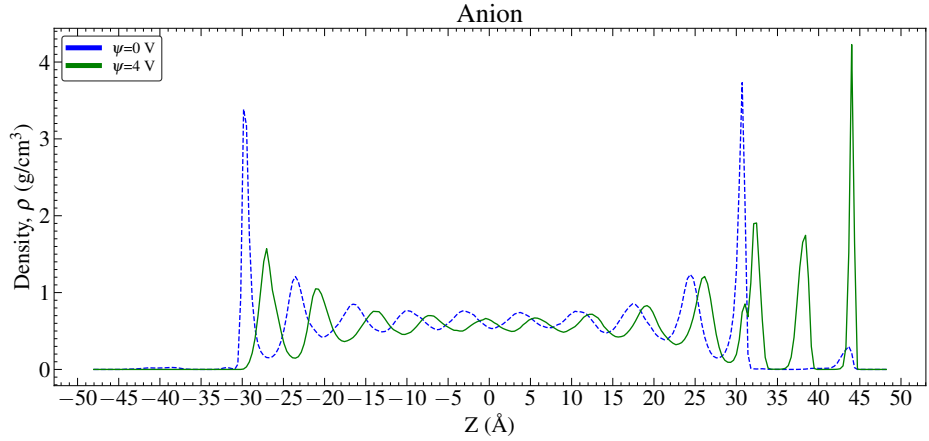

Figure 3: Density profile of anions before and after applying a potential difference in CNT pores.

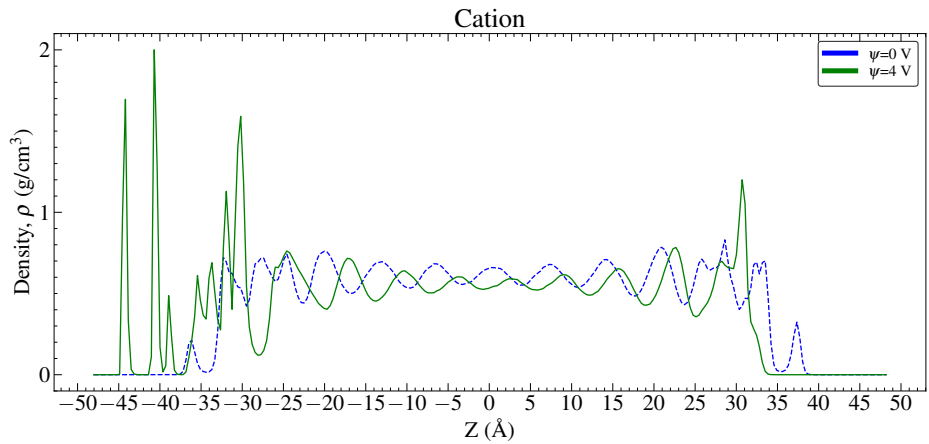

Figure 4: Density profile of cations before and after applying a potential difference in CNT pores.

cation of a potential, a substantial density peak emerges within the anode pore, signifying the entry of cations into the anode pores. Consequently, this approach allows for the complete separation of anions and cations, effectively attaching them to the cathode and anode, respectively.

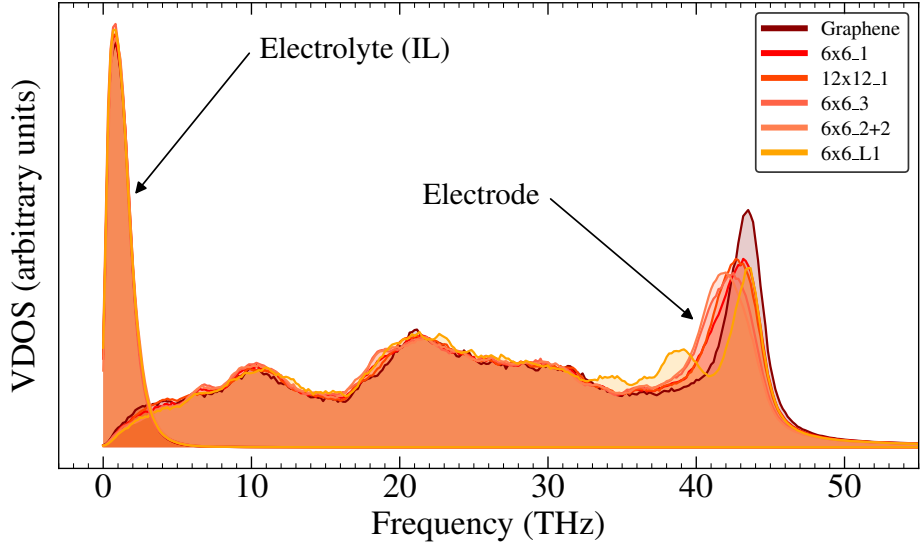

Figure 5: Vibrational density of states (VDOS) of electrodes and IL of a few selected systems.

## VDOS

The vibrational density of states (VDOS) of IL and all the electrodes was investigated by computing the Fourier transform of the velocity autocorrelation function (VACF).[1,2]

$$P(f) = \frac{1}{\sqrt{2\pi}} \int_{-\infty}^{+\infty} C(t).e^{ift} dt, \qquad (1)$$

where $P(f)$ is the VDOS, $f$ is the phonon frequency, $C(t)$ is the VACF given by

$$C(t) = \langle v(t).v(0) \rangle / \langle v(0).v(0) \rangle \qquad (2)$$

where $v(t)$ is the velocity of atom at time $t$.

The VDOS of electrodes and IL has plotted in Fig. 3. The results show that adding CNTs to graphene reduces the peak and slightly shifts the vibrational modes to lower frequencies. However, insignificant spectrum change is observed for lower frequencies and IL. Also, the

5

VDOS overlap $S$, of the electrode and the electrolyte was calculated using the equation,[3]

$$S = \frac{\int_0^\infty P(f)_{Electrode} P(f)_{IL} \, df}{\int_0^\infty P(f)_{Electrode} \, df . \int_0^\infty P(f)_{IL} \, df} \quad (3)$$

The energy carriers in IL are molecules and the vibrational spectral dynamics in bulk liquids cannot be defined. However, under extreme confinements, surface charges, electric fields, and so on, a high ordering of IL (solid-like structure) near the interface is possible. Nevertheless, the change in VDOS overlap for different electrodes was found to be insignificant, indicating the vibrational coupling between IL and the electrode is unaltered with a change in electrode structure. Therefore, the change in interfacial thermal transport with the addition of CNTs is not attributed to the change in the vibrational spectrum as observed in previous works.[4,5]



\end{document}